%% file: paper4.tex
\documentclass[amsmath,amssymb,showkeys,12pt,tightenlines,nobibnotes,nofootinbib]{revtex4}

\usepackage{amsfonts,amssymb,amsmath}
\usepackage{epsfig}
\usepackage{fancyhdr}
\usepackage{color,graphics,pst-grad}

\def\ind{{\mathbb I}}
\newcommand{\ud}{\,\mathrm{d}}
\newcommand{\e}{\mathrm{e}}

\newcommand{\bx}{\mathbf{x}}
\newcommand{\bze}{\mathbf{0}}
\newcommand{\bc}{\mathbf{c}}
\newcommand{\by}{\mathbf{y}}

\renewcommand{\P}{\mathbb{P}}
\newcommand{\N}{\mathcal{N}}

\newcommand{\vers}{\rightarrow}

\begin{document}

\title{Geometrical organization of solutions\\ to random linear
Boolean equations}
%Distance properties in random XORSAT\\
%and the $x$-satisfiability threshold}

\author{Thierry Mora}
\affiliation{Laboratoire de Physique Th\'eorique et Mod\`eles statistiques,\\
UMR~8626, CNRS and Universit\'e Paris Sud\\
Orsay Cedex, F-91405 (France)}

\author{Marc M\'ezard}
\affiliation{Laboratoire de Physique Th\'eorique et Mod\`eles statistiques,\\
UMR~8626, CNRS and Universit\'e Paris Sud\\
Orsay Cedex, F-91405 (France)}

\begin{abstract}
The random XORSAT problem deals with large random 
linear systems of Boolean variables. The difficulty of
such problems is controlled by the ratio of number of equations
to number of variables. It is known that in some range of values of 
this parameter,  the space of solutions  breaks
into many disconnected  clusters. Here we study
precisely the corresponding geometrical organization.
In particular, the distribution of distances
between these clusters is computed by the cavity method.
This allows to study the  `$x$-satisfiability' threshold, the
critical density of equations where there exist two solutions
at a given distance. 

\end{abstract}


\maketitle

\section{Introduction}\label{sec:intro}
%?? Rajouter des refs??
Constraint Satisfaction Networks (CSN) are problems involving many discrete
variables, with values in a finite alphabet, related by low density
constraints: each constraint involves a finite number of variables. This kind
of problems arise in many branches of science, from statistical physics (spin
or structural glasses \cite{MezardParisi87b}) to information theory (low-density parity-check (LDPC) codes \cite{Gallager62,MacKay03}) and combinatorial optimization (satisfiability, colouring \cite{Papadimitriou94}). The
`thermodynamic limit' of such problems is obtained when the number of
variables and the number of constraints go to infinity, keeping their ratio,
the density of constraints $\alpha$, fixed. A lot of attention has been
focused in recent years on the study of random CSN, both because of their
practical interest in coding, and also as a means to study ``typical case''
complexity (as opposed to the traditional worst case complexity analysis).
Many CSN are known to undergo a SAT-UNSAT phase transition when the density of
constraints increases: there is a sharp threshold separating a SAT phase where
all constraints can be satisfied with probability one in the thermodynamic
limit from an UNSAT phase where, with probability one, there is no
configuration of the variables satisfying all the constraints. While the
existence of a sharp threshold has been proved by Friedgut \cite{Friedgut99} for satisfiability
and colouring, there is no yet any rigorous proof of the widely accepted
conjecture according to which the threshold density of constraints converges
to a fixed value $\alpha_c$ in the thermodynamic limit.

Recent years have seen the upsurge of statistical physics methods in the study
of CSN. In particular, the replica method and the cavity method have been used
to study the phase diagram \cite{MezardParisi02,MezardZecchina02,MuletPagnani02}. Their most spectacular results are some arguably
exact (but not yet rigorously proved) expressions for $\alpha_c$, and the
existence of an intermediate SAT phase, in a region of constraint density
$]\alpha_d,\alpha_c[$, where the space of solutions is split into many
clusters, far away from each other. This clustering is an important building
block of the theory: it is at the origin of the necessity to use the cavity
method at the so-called one-step replica symmetry breaking (1RSB) level; this method can be seen as a
message-passing procedure and used as an algorithm for finding a SAT
assignment of the variables. This algorithm, called survey propagation, turns
out to be very powerful in satisfiability and colouring, and its effectiveness
can be seen as one indirect piece of evidence in favour of clustering. On
intuitive grounds, clustering is often held responsible for blocking many
local search algorithms \cite{SemerjianMonasson04}. Although there does not exist any general discussion of this statement, this phenomenon was thoroughly investigated in the case of XORSAT \cite{MontanariSemerjian05-2}.

The clustering effect can be studied in a more formal way by introducing the
notion of $x$-satisfiability \cite{MezardMora05,MoraMezard05}. A CSN with $N$ variables is said $x$-satisfiable
($x$-SAT) if there exists a {\it pair} of SAT assignments of the variables
which differ in a number of variables $\in [Nx-\epsilon(N),Nx+\epsilon(N)]$.
Here $x$ is the reduced distance, which we keep fixed as $N$ goes to
infinity. The resolution $\epsilon(N)$ has to be sub-linear in $N$:
$\lim_{N\rightarrow\infty}\epsilon(N)/N=0$, but its precise form is
unimportant for our large $N$ analysis. For example we can choose
$\epsilon(N)=\sqrt{N}$. For many random CSN, it is reasonable to conjecture, in parallel
with the existence of a satisfiability
threshold, that $x$-satisfiability has a sharp threshold
$\alpha_c(x)$ such that:
\begin{itemize}
\item if $\alpha<\alpha_c(x)$, a random formula is $x$-SAT almost surely.
\item if $\alpha>\alpha_c(x)$, a random  formula  is $x$-UNSAT almost surely.
\end{itemize}
This conjecture has been proposed for $k$-satisfiability of random Boolean formulas where each clause involves exactly $k$ variables with $k\ge 3$. So far
only a weaker conjecture, analogous to  Friedgut's theorem
\cite{Friedgut99}, has been established \cite{MoraMezard05}. It  states the
existence of a non-uniform threshold $\alpha_c^{(N)}(x)$. Rigorous bounds on
$\alpha_c(x)$ have been found in \cite{MoraMezard05} for the $k$-satisfiability
problem with $k\ge 8$, using moment methods developed in \cite{AchlioptasPeres04}, but so far
this $x$-satisfiability threshold has not been computed.

In this paper we compute the $x$-satisfiability threshold $\alpha_c(x)$ in the
random XORSAT problem using the cavity method. This is a problem of random
linear equations with Boolean algebra. It is important because many efficient
error correcting codes are based on low-density parity-checks, the decoding of
which involves precisely such linear systems. It is also one of the best
understood case of CSN. In particular, efforts to extend the replica method
\cite{Monasson98} and the cavity method \cite{MezardParisi01} to deal with
models defined on finite-connectivity lattices, have resulted in the first
exact (but non-rigorous) derivation of its phase diagram \cite{RicciWeigt01}.
Later, a clear characterization of these clusters, combined with simple
combinatoric arguments, gave a rigorous base to these predictions
\cite{DuboisMandler02,CoccoDubois03,MezardRicci03}. These works have computed
the phase diagram in details and provide expressions for the two thresholds
$\alpha_d<\alpha_c<1$.

Our computation of $\alpha_c(x)$ confirms this known structure, and it also
provides insight into the geometrical structure of clusters. We find that
$\alpha_c(x)$ is non monotonic (see fig. \ref{fig:alphaxphasek3}), which
confirms the existence of gaps in distances where there does not exist any
pair of solutions. 

The method used in our computation is in itself interesting. It turns out that
it is not possible to compute $\alpha_c(x)$ directly, by fixing $x$ and
varying $\alpha$. Instead, we work at a fixed value of $\alpha$ and introduce
a probability distribution for pairs of SAT assignments, where the distance
between the solutions plays the role of the energy. The computation of the
entropy as a function of the energy, and more precisely the computation of the energies where it
vanishes, then allows to reconstruct $\alpha_c(x)$. Our computation thus
involves a mixture of hard constraints (the fact that the two assignments
must satisfy the XORSAT formula), and soft constraints (the Boltzmann
weight which depends on their distance). This is reflected in the structure of
the cavity fields that solve this problem.

The remainder of this paper is organized as follows. The next section
 introduces some notations. In section \ref{sec:SP}, we analyse classical
 Survey Propagation on XORSAT and show its equivalence with the ``leaf
 removal'' \cite{MezardRicci03} or ``decimation'' \cite{CoccoDubois03}
 algorithm. This analysis allows to re-derive the phase diagram of XORSAT, and
 sets up useful notations and concepts for later computations. In section
 \ref{sec:single} we perform a statistical mechanics analysis of weight
 properties in a single cluster using the cavity method. Section
 \ref{sec:diameter} applies this formalism to the computation of the cluster
 diameter, while section \ref{sec:inter} is devoted to the evaluation of
 inter-cluster distances.
%A special attention will be devoted to the core, where the distance enumerator function is simpler to evaluate.
In section \ref{sec:conclu} we sum up and discuss our results.

\section{Notations and definitions}

A XORSAT formula is defined on a string of $N$ variables $x_1,x_2,\ldots,x_N \in \{0,1\}$ by a set of $M$ parity checks of the form:
\begin{equation}\label{eq:defxorsat}
\sum_{i\in V(a)} x_i = y_a\ (\mathrm{mod}\ 2),\quad\textrm{for all }a=1,\ldots,M
\end{equation}
where $y_a\in\{0,1\}$.
Here $V(a)\subset \{1,\ldots,N\}$ is the subset of variables involved in parity check $a$. Later on $i\in a$ shall be used as a shorthand for $i\in V(a)$.

Eq.~\eqref{eq:defxorsat} can be rewritten in the matricial form:
\begin{equation}\label{eq:defmatrix}
A\bx=\by\ (\mathrm{mod}\ 2),\qquad A=\{A_{ia}\}_{i\in [N],\,a\in [M]}
\end{equation}
where $A_{ia}=1$ if $i\in a$ and $A_{ia}=0$ otherwise. The pair $F=(A,\by)$
defines the formula. Such a linear system can be solved in polynomial time by
Gaussian elimination. If a formula has solutions, it is SAT; otherwise, it is
UNSAT. The thermodynamics limit is $N\to \infty, \; M\to\infty$ with a fixed
density of constraints $\alpha=M/N$.

In this paper we specialize to random $k$-XORSAT formul{\ae}, where each
equation involves a subset of $k$ variables, chosen independently with uniform
probability among the $\binom{N}{k}$ possible ones, and each $y_a$
independently takes value $0$ or $1$ with probability $1/2$. 
One important characterization of a XORSAT formula $F=(A,\by)$ is the
number  $\N_N(F)$ of assignments of the Boolean variables $\bx$ which satisfy
all the equations, and the corresponding entropy density
\begin{equation}
s_N(F)=\frac{1}{N}\log \N_N(F)
\end{equation}
Logarithm is base 2 throughout the paper.
 Using a spin
representation $\sigma_i=(-1)^{x_i}$, the $k$-XORSAT problem can also be
mapped onto a spin glass model where interactions involve products of $k$
spins (the variables $(-1)^{y_a}$ then play the role of quenched random
exchange couplings) \cite{RicciWeigt01}, and the question of whether a formula is SAT is
equivalent to asking whether the corresponding spin-glass instance is frustrated.

Previous work \cite{RicciWeigt01,DuboisMandler02,CoccoDubois03,MezardRicci03} has shown that:
\begin{itemize}
\item for $\alpha<\alpha_d(k)$, the formula is SAT, almost surely (i.e. with
probability $\vers 1$ as $N\vers\infty$). The solution set forms one big
connected component, the entropy density concentrates at large $N$ to
$(N-M)/N=1-\alpha$ ; This phase is called the EASY-SAT phase.
\item for $\alpha_d(k)<\alpha<\alpha_c(k)$, the formula is still SAT almost
surely, but the solution set is made of an exponentially large  (in $N$)  number of
components far away from each other (in the following we shall give a precise definition
of these clusters); The entropy density also concentrates at large $N$ to
$(N-M)/N=1-\alpha$. This is the HARD-SAT phase.
\item for $\alpha>\alpha_c(k)$ (with $\alpha_c(k)<1$), the formula is UNSAT
almost surely. The entropy is $-\infty$. This second transition is the usual SAT-UNSAT transition.
\end{itemize}

The fact that, throughout the SAT phase ($\alpha<\alpha_c(k)$), the
entropy density concentrates to $1-\alpha$ is not surprising: it can be
understood as the fact that matrix $A$ has rank $M$ almost surely in the SAT
phase. The intuitive reason is that, each time there exists a linearly
dependent set of checks, the choice of $y_a$ has probability $1/2$ to lead to
a contradiction. So the rank of $A$ cannot differ much from $M$ in the SAT
phase. From the point of view of linear algebra, the existence of the
clustered phase, i.e. the fact that the vector subspace of SAT assignments
breaks into disconnected pieces, is more surprising, as is the discontinuity
of $ s_N(F)$ at the transition $\alpha_c$. These two aspects are in fact
related: the quantity which vanishes at the SAT-UNSAT transition is actually
the log of the number of clusters of solutions, while each cluster keeps a
finite volume.

We will study the geometric properties of the space of 
solutions for random $k$-XORSAT in the 
HARD-SAT phase using the notion of $x$-satisfiability. In terms of solutions 
of linear equations, we want to know if there exist two Boolean vectors $\bx$ 
and $\bx'$ which both satisfy $A\bx=A\bx'=\by$, where the Hamming distance $d_{\bx,\bx'}\equiv (\bx-\bx')^2=Nx$.
Clearly, if such a pair exists, $\bx-\bx'$ is solution to the homogeneous (`ferromagnetic') problem
where $\by=\mathbf{0}$:
\begin{equation}
A(\bx-\bx')=\mathbf{0}
\end{equation}
Therefore, a formula $F=(A,\by)$ is $x$-SAT if and only if $F$ is SAT and if
there exists a solution $\bx$ to the homogeneous system $A\bx=\mathbf{0}$ of
{weight} $d_{\bx,\mathbf{0}}\approx Nx$ (the {\it weight} is by definition the
distance to $\bze$).
Note that for $x=0$, this second condition is automatically 
fulfilled, and $x$-satisfiability is equivalent to satisfiability.
This linear space structure also implies that the set of solutions looks the same
seen from any solution in the SAT phase: the number of solutions at distance $d$ of any given solution $\bx_0$ is independent from $\bx_0$.

Distance properties can also be investigated directly by evaluating extremal
distances between solutions.
To that end we define three distances: (a) the
cluster diameter $d_1$, i.e. the largest Hamming distance between solutions
belonging to the same cluster;  this diameter is independent 
of the cluster; (b) the minimal
and maximal inter-cluster distances $d_2$ and $d_3$, i.e. the smallest (resp.
largest) Hamming distance between solutions belonging to distinct clusters.
All three distances are assumed to be self-averaging in the thermodynamic
limit of the random problem: $x_1(\alpha)=d_1/N$, $x_2(\alpha)=d_2/N$ and $x_3(\alpha)=d_3/N$ shall
denote the corresponding limits. In the particular case where $k$ is even, the
formula is invariant under the transformation $\bx\leftrightarrow
\bx+\mathbf{1}\ (\textrm{mod}\ 2)$, which is reflected in terms of distances
by a symmetry with respect to $x=1/2$: $x\leftrightarrow 1-x$. A direct
consequence is that $x_3(\alpha)=1-x_2(\alpha)$, and that a fourth weight,
defined as $1-x_1(\alpha)$, will also come into play. These distance functions
are related to  the $x$-satisfiability threshold
as follows: at fixed $\alpha$, a formula is $x$-SAT almost surely iff
\begin{itemize}
\item $x\in [0,x_1(\alpha)]\cup [x_2(\alpha),x_3(\alpha)]$ when $k$ is odd.
\item $x\in [0,x_1(\alpha)]\cup [x_2(\alpha),1-x_2(\alpha)]\cup [1-x_1(\alpha),1]$ when $k$ is even.
\end{itemize}

We will now compute $x_1,x_2,x_3$ with the cavity method.
%Remarkably, the ferromagnetic version of the XORSAT problem ($y_a=0$ for all
%$a$) shares many common features with its random counterpart. If one puts
%aside the trivial solution ($x_i=0$ for all $i$), the system undergoes the
%same two transitions with the same behaviour: in physical terms, the liquid
%and glassy phases ``ignore'' the crystal corresponding to the trivial solution
%$\bx=\bze$\cite{FranzMezardRicciWeigtZecchina}. All properties described in the
%remainder of this paper also apply to ferromagnetic XORSAT.

\section{Leaf removal as an instance of Survey Propagation}\label{sec:SP}
XORSAT formul{\ae} are conveniently represented by factor graphs, called {\em
Tanner graphs}, in which variables and checks form two distinct types
of nodes, with the simple rule that the edge $(i,a)$ between $i$ and $a$ is
present if $i\in a$.

An example of a Tanner graph and its associated linear system is shown below:

\bigskip

\begin{minipage}{.7\linewidth}
\begin{equation}\nonumber
\begin{split}
(a)\quad & {x_1}+{x_2}+{x_{3}}=0\quad (\mathrm{mod}\ 2)\\
(b)\quad & {x_2}+{x_3}=1\quad (\mathrm{mod}\ 2) \\
(c)\quad & {x_2}+{x_3}+{x_4}=1\quad (\mathrm{mod}\ 2)
%\\& \cdots 
\end{split}
\end{equation}
\end{minipage}
\begin{minipage}{.12\linewidth}
\begin{center}
\epsfig{file=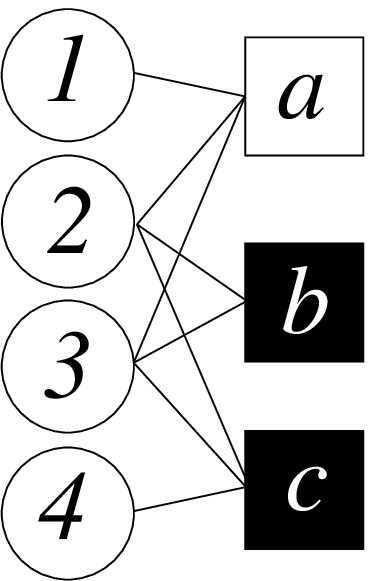,width=.9\linewidth}
\end{center}
\end{minipage}

\bigskip

The number of variables involved in a check $a$, denoted by $|V(a)|$, is the
degree of $a$ in the factor graph. Here we study $k$-XORSAT where this degree is fixed to $k$. Similarly, if $V(i)$ denotes the set
of parity checks in which $i$ is represented, $|V(i)|$ is the degree of
$i$ in the factor graph. The degrees of checks are commonly referred to as {\em
right-degrees}, and those of variables as {\em left-degrees}.
 The infinite-length (thermodynamic)
limit is obtained by sending $N$ and $M$ to infinity while keeping the ratio
$\alpha=M/N$ fixed. In this limit, the distribution of left-degrees is a
Poisson law of parameter $k\alpha$: The probability of a variable having
degree $\ell$ is $\pi_{k\alpha}(\ell)$, where
$\pi_x(\ell)=\exp(-x)x^\ell/\ell!$.

Here we use the {\em leaf removal algorithm} (LR) in order
to obtain a precise definition of the notion of ``cluster'' or ``component'' of
solutions, one which is valid also for finite $N$. 
The algorithm proceeds as follows: pick a variable of degree one (called a
{\it leaf}), remove it as well as the only check it is connected to. Continue
the process until there remains no leaf. The interest of this algorithm is
easily seen: a variable on a leaf can always be assigned in such a way that the
(unique) check to which it is connected is satisfied.

The linear system remaining after leaf removal is independent of the order in
which leaves are removed. It is called the {\it core}. A `core check' is a check which only involves core variables. If the core is empty, the
problem is trivially SAT. In general, given a solution of the core, one can
easily reconstruct a solution of the complete formula by running leaf removal
in the reverse direction, in a scheme which we refer to as {\em leaf
reconstruction}. In this procedure, checks are added one by one along with
their leaves, starting from the core. If an added check involves only one
leaf, the value of that variable is determined uniquely so that the check is
satisfied. If the number of leaves $k'$ is greater that $1$, one can choose
the joint value of those leaves among $2^{k'-1}$ possibilities. The process is
iterated until the complete factor graph has been rebuilt. Given a core
solution, one can construct many solutions to the complete formula. Variables
which are uniquely determined by the core solution are called {\it frozen},
and variables that can fluctuate are called {\it floppy}. Of course, by
definition, the frozen part includes the core itself. A core solution defines
a {\it cluster}. All solutions built from the same core solution belong to the
same cluster. We shall see later how this definition fits in the intuitive
picture that we sketched previously in terms of connectedness.

We propose here an alternative to the leaf removal algorithm, which also
builds the core, but keeps actually more information. The approach is inspired
by the cavity method, and is a special instance of Survey Propagation (SP)
\cite{MezardZecchina02}. To each edge $(i,a)$, one assigns two numbers $\hat
m^t_{a\vers i}$ and $m^t_{i\vers a}$ belonging to $\{0,1\}$, updated as
follows:

\begin{itemize}
\item At $t=0$, $\hat m^0_{a\vers i}=1$, $m^0_{i\vers a}=1$ for all edges $(i,a)$.
\item $m^{t+1}_{i\vers a} = 1 -\prod_{b\in i-a} (1-\hat m^t_{b\vers i})$.
\item $\hat m^t_{a\vers i}=\prod_{j\in a-i} m^t_{j\vers a}$.
\item Stop when $\hat m^{t+1}_{a\vers i}=\hat m^{t}_{a\vers i}$ for all $(i,a)$,
\end{itemize}
Here $a\in i$ is a shorthand for $a\in V(i)$.

The interpretation of $m^t_{i\vers a}=1$ is: ``variable $i$ is constrained at
time $t$ in the absence of check $a$'', and $\hat m^t_{a\vers i}=1$: ``check
$a$ constrains variable $i$ at time $t$''. One also defines $M^t_i=1
-\prod_{a\in i} (1-\hat m^t_{a\vers i})\in\{0,1\}$. This number indicates
whether node $i$ is constrained at time $t$ ($M^t_i=1$) or not ($M^t_i=0$).

At $t=0$, all variables are constrained. The algorithm consists in detecting
the under-constrained variables, and propagating the information through the
graph to simplify the formula. At the first step, only variables of degree one
are affected: if $i$ is of degree one and is connected to $a$, $m^1_{i\vers
a}=1-\prod_{\emptyset}=0$. This, in turn, gives freedom to $a$, which no
longer constrains its other variables: $\hat m^1_{a\vers j}=0$, for $j\in
a-i$. This effectively removes $a$ and $i$ from the formula, just as in the
leaf removal algorithm. In the subsequent steps of the iteration, will be
considered as a {\em leaf} (in the LR sense), a variable $i$ such that there
exists exactly one $a\in i$ such that $\hat m^t_{a\vers i}=1$. In that case we
have $m^{t+1}_{i\vers a}=0$, thus implementing a step of LR.

Let us add a  word about the
term ``Survey Propagation'' we have used so far. Analysis of the 1RSB cavity
equations at zero temperature \cite{MezardRicci03} (see
\cite{MezardZecchina02} for a more complete discussion in the case of $k$-SAT)
shows that cavity biases fall into two categories, depending on the edge we
consider: either a warning is sent (compelling to take value $0$ or $1$
depending on the cluster, with probability one half for each), or no warning
is sent. 
(In more technical terms, the survey propagation reduces to warning propagation). %ICI
The first situation corresponds in our language to $\hat m_{a\vers
i}=1$ and the second to $\hat m_{a\vers i}=0$. Similarly, we have $m_{i\vers
a}=1$ if the cavity field is non-zero, and $m_{i\vers a}=0$ otherwise.
Therefore our algorithm carries the same information as Survey Propagation.

The interest of SP over leaf removal is that it keeps track of the leaves
which are uniquely determined by their check. For example, if two or more
leaves are connected to the same check $a$ at time $t$, at time $t+1$ one has
$\hat m^{t+1}_{a\vers i}=0$ for all $i\in a$, reflecting the fact that $a$
cannot uniquely determine the value of several leaves. Conversely, if $a$ is
connected to a unique leaf $i$ and if one has: $m^t_{j\vers a}=1$ for all
$j\in a-i$, then one gets $\hat m^t_{a\vers i}=1$, reflecting the fact that,
the variables $\{x_j\}_{j\in a-i}$ being fixed in the absence of $a$, $i$ is
determined uniquely.

A little reasoning shows that when the algorithm stops ($t=t_f$), $i$ is
frozen iff $M^{t_f}_i=1$, and $i$ belongs to the core iff there exists at
least two checks $a,b\in i$ such that $\hat m^{t_f}_{a\vers i}=\hat
m^{t_f}_{b\vers i}=1$. In the final state, we say that the directed edge
$i\vers a$ is frozen if $m_{i\vers a}\equiv m^{t_f}_{i\vers a}=1$, and that
$a\vers i$ is frozen if $\hat m_{a\vers i}\equiv \hat m^{t_f}_{a\vers i}=1$.
In the opposite case, edges are called floppy.
This version of SP is strictly equivalent to the {\it Belief
Propagation} algorithm used for decoding Low-Density Parity-Check codes on the
binary erasure channel, also called ``Peeling decoder'' in that
context.

\begin{figure}
\begin{center}
\epsfig{file=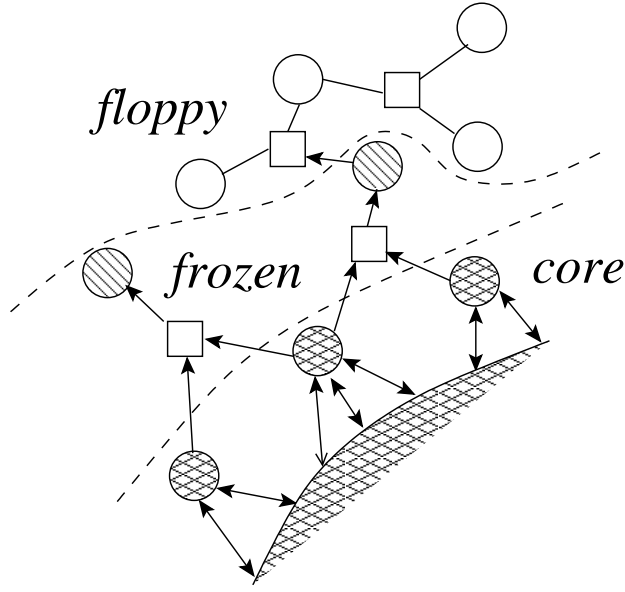,width=.5\linewidth}
\caption{\label{fig:leafrem} A example of a fixed point of SP. Circles represent variable nodes, and squares check nodes. An arrow means that message $m$ or $\hat m$ has value $1$, that is, that the directed edge is frozen when SP stops. Leaf removal propagates null messages from the outer leaves down to the core, while ``leaf reconstruction'' propagates non-null messages from the core up the frozen part.
}
\end{center}
\end{figure}

SP can be studied by density evolution in order to derive the phase diagram,
as in \cite{MezardRicci03}. Let us briefly survey this study for completeness.
 The statistics of messages at time $t$
is described by two numbers:
\begin{equation}
v^t=\frac{1}{Mk}\sum_{(i,a)} \delta(m^t_{i\vers a},0)\ \ \ , \ \ \ 
w^t=\frac{1}{Mk}\sum_{(i,a)} \delta(\hat m^t_{a\vers i},0)\ ,
\end{equation}
where the sums run over all edges of the Tanner graph.
When $N\vers\infty$, these densities are governed by evolution equations:
\begin{equation}\label{eq:de}
\begin{split}
v^{t+1}&=\sum_{\ell} \pi_{k\alpha}(\ell) (w^t)^\ell= \exp\left[-k\alpha(1-w^t)\right]\\
w^t&=1-{\left[1-v^t\right]}^{k-1}\ ,
\end{split}
\end{equation} which are initialized with $v^0=w^0=0$.
These equations are exact if the Tanner graph is a tree. In our case
the graph is locally tree-like (it is a tree up to finite distance when seen from
a generic point), and one could set up a rigorous proof of \eqref{eq:de}
using the methods developed in \cite{RichardsonUrbanke06}. %ICI
 
The fixed point of these equations is given by the {\em cavity equation}:
\begin{equation}\label{eq:cavityenergetic1rsb}
w=1-{\left\{1-\e^{-k\alpha(1-w)}\right\}}^{k-1}.
\end{equation}
Setting $\lambda=k\alpha(1-w)$, Eq.~\eqref{eq:cavityenergetic1rsb} can be rewritten as:
\begin{equation}
\lambda=k\alpha(1-\e^{-\lambda})^{k-1}
\end{equation}

%\begin{figure} ICI
%\begin{center}
%
%\input{devo1.tex}
%
%\input{devo2.tex}
%
%\caption{\label{fig:devo}Density evolution for $\alpha>\alpha_d$ (top) and $\alpha<\alpha_d$ (bottom).}
%\end{center}
%\end{figure}

When $\alpha<\alpha_d$, the unique fixed point is $\lambda=0$ (i.e. $w=1$).
This  means that  the core is empty. For $\alpha>\alpha_d$ however, there remains an extensive core 
%(see Fig.~\ref{fig:devo}) 
of size
\begin{equation}
N_c=N\left[\sum_{\ell\geq 2}\pi_{k\alpha}(\ell) (1-w^{\ell}-\ell w^{\ell-1})\right]=N\left[1-(1+\lambda)\e^{-\lambda}\right]
\end{equation}
while the number of frozen variables is
\begin{equation}
N_f=N\left[\sum_{\ell\geq 2}\pi_{k\alpha}(\ell) (1-w^{\ell})\right]=N\left[1-\e^{-\lambda}\right]
\end{equation}
 The number of core checks is:
\begin{equation}
M_c=M(1-v)^k=\alpha N\left[1-\e^{-\lambda}\right]^k.
\end{equation}

The left-degree distribution (with respect to core checks) inside the core is given by a truncated Poissonnian:
\begin{equation}\label{degreedistcore}
P_c(\ell)=\frac{1}{e^\lambda-1-\lambda} 
\frac{\lambda^\ell}{\ell!}\ \ind(\ell\geq 2)\ ,
\end{equation}
where $\ind$ is the indicator function.

One can show that the leaf removal algorithm conserves the uniformity of the
ensemble. Therefore, the core formula is a random XORSAT formula with
right-degree $k$ and left-degree distribution $P_c(\ell)$ given by
\eqref{degreedistcore}. The number of solution to such a formula is known to
concentrate to its mean value when the size goes to infinity
\cite{MezardRicci03,DuboisMandler02}. In the case of the core formula, this
number is simply $2^{N_c-M_c}$ if $N_c\geq M_c$, and $0$ otherwise. Recalling
that the complete formula has solutions if and only if the core formula does,
we find that the SAT-UNSAT threshold $\alpha_c$ is given by the equation:
\begin{equation}
1-(1+\lambda)\e^{-\lambda}=\alpha\left[1-\e^{-\lambda}\right]^k.
\end{equation}
The number of clusters is characterized by the {\it complexity} or
{\it configurational entropy}, that is the logarithm of the number of core solutions:
\begin{equation}\label{eq:complexity}
\Sigma(\alpha)=\frac{1}{N}\log(\textrm{\# clusters})=\frac{N_c-M_c}{N}=1-(1+\lambda)\e^{-\lambda}-\alpha\left[1-\e^{-\lambda}\right]^k
\end{equation}
We recall  that the group structure of the solution set implies that all clusters have the same internal structure. Their common internal entropy is therefore given by:
\begin{equation}
s_{\rm inter}=1-\alpha - \Sigma(\alpha)
\end{equation}
where we have used the fact that the total entropy is $1-\alpha$.

Let us comment on the relationship between our definition of clusters
and the more traditional one. Usually, clusters
are defined as the ``connected'' components of the solution set, where
connectedness is to be understood in the following way: two solutions are
connected if one can go from one to the other by a sequence of solutions
separated by a finite Hamming distance (when $N\vers\infty$). To make contact
with our own definition of clusters, one needs to prove two things. First,
that two solutions built from the same core solution are connected. Second,
that two core solutions are necessarily separated by an extensive Hamming
distance ($\geq c N$, with $c$ constant), which implies that two solutions
built from two distinct core solutions are not connected. Both proves can be
found in \cite{MezardRicci03}. This reconciles our definition (which holds for
any single instance of XORSAT) with the usual one (which only makes sense for
infinite-length ensembles).

\section{Distance landscape: thermodynamical approach}\label{sec:single}

As we have already observed, studying pairs of solutions is equivalent to studying solutions to the ferromagnetic problem. Indeed, if $S$ denotes the affine subspace of solutions to $A\bx=\by$, and $S_0$ the vector subspace of solutions to $A\bx=\bze$, we have:
\begin{equation}
S\times S=\{(\bx',\bx'+\bx),(\bx',\bx) \in S\times S_0\}
\end{equation}
In particular, distances in $S$ are reflected by weights in $S_0$. Therefore, in order to study the range of attainable distances between solutions, one just needs to study the range of possible weights in $S_0$. To that end we set a thermodynamical framework in which the weight plays the role of an energy:
\begin{equation}
E(\bx)\equiv |\bx|=\sum_i \delta_{x_i,1}.
\end{equation}
The Boltzmann measure at temperature $\beta^{-1}$ is thus defined by:
\begin{equation}\label{eq:boltzmann}
\P(\bx,\beta)= \frac{1}{Z(\beta)}\prod_a\delta_{\mathbb{F}_2}\left(\sum_{i\in a}x_i,0\right)2^{-\beta|\bx|}
\end{equation}
where the normalization constant $Z(\beta)$ is the partition function. The Dirac delta-function, here defined on the two-element field $\mathbb{F}_2$, enforces that only configurations of $S_0$ are considered.
Remarkably, this measure is formally similar to the one used to infer the most probable codeword under maximum-likelihood decoding in Low-Density Parity-Check (LDPC) codes on the Binary Symmetric Channel \cite{NishimoriBook01}. In fact, as we shall see soon, some of the methods used to solve both problems share common aspects.

A very useful scheme for estimating marginal probabilities in models defined on sparse graphs is the cavity method \cite{MezardParisi01}, which we have already mentioned in the previous section. Let $p^{x}_{i\vers a}$ be the probability that $x_i=x$ under the measure defined by \eqref{eq:boltzmann}, where the link $(i,a)$ has been removed. The replica symmetric (RS) cavity method consists in computing the cavity marginals $p^x_{i\vers a}$ (viewed as variable-to-check messages) using a closed set of equations where check-to-variable messages are also introduced as intermediate quantities. These second-kind messages are denoted by $q^x_{a\vers i}$ and are proportional to the probability that $x_i=x$ when $i$ is connected to $a$ only. Messages are updated until convergence with the following rules:
\begin{eqnarray}\label{eq:cavity1}
p^{x_i}_{i\vers a}&=&\frac{1}{Z_{i\vers a}}\prod_{b\in i-a}q^{x_i}_{b\vers i}2^{-\beta\delta_{x_i,1}}\\
q^{x_i}_{a\vers i}&=&\sum_{\{x_j\}_{j\in a-i}} \prod_{j\in a-i} p^{x_j}_{j\vers a}\ \delta_{\mathbb{F}_2}\left(\sum_{j\in a}x_j,0\right)\label{eq:cavity2}
\end{eqnarray}
where $Z_{i\vers a}$ is a normalization
constant. When convergence is reached, marginal probabilities are obtained as:
\begin{equation}
p^{x_i}_i\equiv \sum_{\{x_j\}_{j\neq i}}\P(\bx,\beta)=\frac{1}{Z_{i+a\in i}}\prod_{a\in i}q^{x_i}_{a\vers i}2^{-\beta\delta_{x_i,1}}
\end{equation}
where $Z_{i+a\in i}$ is also a normalization constant.
Continuing the analogy with codes, it is interesting to note that these cavity equations are
identical \cite{YedidiaFreeman02} to
the Belief Propagation (BP) equations \cite{KschischangFrey01} used to decode 
messages with LDPC codes on the Binary Symmetric Channel.

It turns out that cavity equations \eqref{eq:cavity1}, \eqref{eq:cavity2} do
not admit a unique solution, as one would expect if the system were replica
symmetric. Instead, let us show that they admit exactly one solution for each
cluster. In a given cluster denoted by $\bc$, let us denote by $c_i$  the value of a frozen variable $i$.  
There exists a solution 
to \eqref{eq:cavity1}, \eqref{eq:cavity2}, where, for every frozen variable
$i$:
\begin{equation}\label{eq:fixing}
\begin{split}
p^x_{i\vers a}&=\delta_{x,c_i}\quad\textrm{if }i\vers a\textrm{ frozen}\\
q^x_{a\vers i}&=\delta_{x,c_i}\quad\textrm{if }a\vers i\textrm{ frozen},
\end{split}
\end{equation}

%ICI
 In order
to show that this is a solution, let us use the SP messages, which provide information on how the
fixing of the core solution forces the values of frozen variables.
For example $m_{i\vers a}=1$ indicates that $x_i$ is entirely determined by
the core solution, supposing that the edge $(i,a)$ has been removed. Consider the  SP fixed point relations
\begin{equation}
\begin{split}
\hat m_{a\vers i}&=\prod_{j\in a-i} m_{j\vers a},\\
m_{i\vers a} &= 1 -\prod_{b\in i-a} (1-\hat m_{b\vers i}) \ .
\end{split}
\end{equation}
They are in fact {\em contained} in the cavity equations
Eqs.~\eqref{eq:cavity1}, \eqref{eq:cavity2}. In fact, the iteration of cavity
equations allows to identify the frozen edges, irrespectively of the cluster
the system falls into.

But the cavity equations also contain `fluctuating' messages, where $p^x$ and $q^x$ are in $]0,1[$, 
which  are {\em de facto} restricted to the floppy part. We parametrize them by  the cavity fields and biases:
\begin{equation}
\beta h^\bc_{i\vers a}=\log\frac{p^{0}_{i\vers a}}{p^{1}_{i\vers a}},
\qquad \beta u^\bc_{a\vers i}=\log\frac{q^{0}_{a\vers i}}{q^{1}_{a\vers i}}
\end{equation}
which satisfy the equations:
\begin{eqnarray}\label{eq:cavintramax1}
h^\bc_{i\vers a}&=&\sum_{b\in i-a} u^\bc_{b\vers i} +1\quad\textrm{with }i\vers a\textrm{ floppy},\\
\beta u^\bc_{a\vers i}&=&2\ \mathrm{atanh}\left[\prod_{j\in a^{nf}-i}\tanh (\beta h^\bc_{j\vers a}/2)\prod_{j\in a^{f}-i}(-1)^{c_j}\right]\quad\textrm{with }a\vers i\textrm{ floppy}\label{eq:cavintramax2}
\end{eqnarray}Note that cavity messages $h^\bc_{i\vers a}$ and $u^\bc_{a\vers i}$ now depend explicitly on the considered cluster, and are uniquely determined by it.
%ICI Unicity?

The multiplicity of solutions to RS cavity equations is a clear sign that the
replica symmetry is broken. The main lesson from this discussion is that
solutions can fluctuate according to two hierarchical levels of statistics:
the first level deals with fluctuations inside a single cluster, i.e.
fluctuations on the floppy part, while the second level deals with the choice
of the cluster. The reduced cavity equations \eqref{eq:cavintramax1},
\eqref{eq:cavintramax2} correctly describe the first level\footnote{Although
the RS Ansatz is unable to describe the whole system, it can reasonably be
assumed to be valid on a single cluster.}, when the system is forced to live
in cluster $\bc$. This leads to defining a new probability measure and
partition function, restricted to $\bc$:
\begin{equation}\label{eq:Zc}
Z_\mathbf{c}(\beta)= \sum_{\bx\in\mathbf{c}}
2^{-\beta\sum_{i=1}^N\delta_{x_i,1}}
\end{equation}
By construction, this system is characterized by the fixing of the frozen edges
\eqref{eq:fixing} and by the reduced cavity equations \eqref{eq:cavintramax1},
\eqref{eq:cavintramax2}. The second level of statistics, i.e. the statistics
over the clusters, is appropriately handled by an 1RSB calculation, and will
be the object of section \ref{sec:inter}. We first focus on the properties of
single clusters under the measure defined by \eqref{eq:Zc}.

The cavity method comes with a technique to estimate the log of the
partition functions, also called potential in our case:
\begin{eqnarray}
\phi(\beta)=-\frac{1}{N}\log Z(\beta)
%\ \ \ ; \ \ \ 
%\phi_\bc(\beta)=-\frac{1}{N}\log Z_\bc(\beta)
\end{eqnarray}
(Note that this quantity differs from the usual free energy by a factor $\beta$). It can be computed within the RS Ansatz by the Bethe formula \cite{YedidiaFreeman02}:
\begin{equation}\label{eq:Bethe}
N\phi(\beta)=\sum_i \Delta \phi_{i+a\in i}-(k-1)\sum_{a} \Delta \phi_a
\end{equation}
where
\begin{equation}\label{eq:phicmax1}
\begin{split}
\Delta \phi_{i+a\in i}&=-\log Z_{i+a\in i}=-\log \sum_{x_i}\prod_{a\in i}q^x_{a\vers i}2^{-\beta\delta_{x_i,1}}\\
\Delta \phi_a&=-\log \sum_{\{x_i\}_{i\in a}} \prod_{i\in a} p^{x_i}_{i\vers a}\ \delta_{\mathbb{F}_2}\left(\sum_{i\in a}x_j,0\right)
\end{split}
\end{equation}
This formula has a rather simple interpretation: $\Delta \phi_{i+a\in i}$ is the contribution of $i$ and its adjacent checks to the potential. When these contributions are summed, each check is counted $k$ times, whence the need to subtract $k-1$ times the contribution of each check $\Delta \phi_a$. Also note that this expression is variational: it is stationary in the messages $\{p_{i\vers a}\}$ as soon as the cavity equations \eqref{eq:cavity1}, \eqref{eq:cavity2} are satisfied.

The RS Ansatz is valid in a single cluster. The single cluster potential
$\phi_\bc(\beta)=-\frac{1}{N}\log Z_\bc(\beta)$ can therefore be computed by
plugging Eqs.~\eqref{eq:fixing}, \eqref{eq:cavintramax1} and
\eqref{eq:cavintramax2} into the Bethe formula \eqref{eq:phicmax1}, provided
one uses the messages corresponding to one given cluster $\bc$. When one is
restricted to a single cluster $\bc$, the range of possible weights is
$[x_\bc,X_\bc]$. The  minimal and
maximal weights can be obtained by sending 
$\beta\vers\pm\infty$. For $\beta\vers\infty$, the second cavity
equation \eqref{eq:cavintramax2} simplifies to:
\begin{equation}
u^\bc_{a\vers i}=\mathcal{S}\left(\prod_{j\in a^{nf}-i} h^\bc_{j\vers a}\prod_{j\in a^{f}-i}(-1)^{c_j}\right)\ \min_{{j\in a^{nf}-i}}|h^\bc_{j\vers a}|\qquad\textrm{with }a\vers i\textrm{ floppy}
\end{equation}
where $\mathcal{S}(x)=1$ if $x>0$, $-1$ if $x<0$ and $0$ if $x=0$.

The ``ground state energy'', i.e. the minimal weight in $\bc$, is obtained as:
\begin{equation}
x_\bc=\lim_{\beta\vers\infty}\partial_\beta \phi_\bc(\beta)=\frac{1}{N}\sum_{i\ {\rm  floppy}}^N \frac{1-\mathcal{S}\left(\sum_{a\in i}u^\bc_{a\vers i}+1\right)}{2}+\frac{1}{N}\sum_{i\ {\rm  frozen}} \delta_{c_i,1}
\end{equation}
The $\beta\vers -\infty$ limit yields very similar equations. These equations will
be analyzed in the next section.

Let us also write down the equations giving the potential,
which will be used in sect.\ref{sec:inter}.   
\begin{equation}
N\phi_\bc(\beta)=\sum_i \Delta \phi^\bc_{i+a\in i}-(k-1)\sum_{a} \Delta \phi^\bc_a
\end{equation}
\begin{equation}
\lim_{\beta\vers\infty}\frac{1}{\beta}\Delta \phi^\bc_{i+a\in i}\equiv \Delta x^\bc_{i+a\in i},\qquad \lim_{\beta\vers\infty}\frac{1}{\beta}\Delta \phi^\bc_a \equiv\Delta x^\bc_a\qquad\textrm{with}
\end{equation}
\begin{eqnarray}
\Delta x^\bc_{i+a\in i}&=&\frac{1}{2}\left(\sum_{a\in i}|u^\bc_{a\vers i}|+1-|\sum_{a\in i}u^\bc_{a\vers i}+1|\right)\quad\textrm{if }i\textrm{ is floppy}\label{eq:potcontrib1}\\
\Delta x^\bc_{i+a\in i}&=&\sum_{a\in i^{nf}} |u^\bc_{a\vers i}|\vartheta(-u^\bc_{a\vers i})\quad\textrm{if }i\textrm{ is frozen and }c_i=0\label{eq:potcontrib2}\\
\Delta x^\bc_{i+a\in i}&=&1+\sum_{a\in i^{nf}} |u^\bc_{a\vers i}|\vartheta(u^\bc_{a\vers i})\quad\textrm{if }i\textrm{ is frozen and }c_i=1\label{eq:potcontrib3}\\
\Delta x^\bc_a&=&\vartheta\left(-\prod_{i\in a^{nf}} h^\bc_{i\vers a}\prod_{i\in a^{f}}(-1)^{c_i}\right)\ \min_{{i\in a^{nf}}}|h^\bc_{i\vers a}|\label{eq:potcontrib4}
\end{eqnarray}

\section{Diameter} \label{sec:diameter}

With our formalism, computing the cluster diameter boils down to computing the maximal weight in cluster $\bze$ (the cluster containing $\bze$). The relevant partition function for this task is:
\begin{equation}\label{eq:Zintra}
Z_\bze(\beta)=2^{-N\phi_\bze(\beta)}=\sum_{\bx\in \bze} \delta_{\mathbb{F}_2}\left(\sum_{i\in a}x_i,0\right)2^{-\beta\sum_{i=1}^N\delta_{x_i,1}}
\end{equation}
When $\beta\vers -\infty$, the solution of the cavity equations corresponding to cluster $\bze$ is characterized by:
\begin{equation}
\begin{split}
p^x_{i\vers a}&=\delta_{x,0}\quad\textrm{if }i\vers a\textrm{ frozen},\\
q^x_{a\vers i}&=\delta_{x,0}\quad\textrm{if }a\vers i\textrm{ frozen},\\
h_{i\vers a}&=\sum_{b\in i-a} u_{b\vers i} +1\quad\textrm{if }i\vers a\textrm{ floppy},\\
u_{a\vers i}&=-\mathcal{S}\left[\prod_{j\in a^{nf}-i} (- h_{j\vers a})\right]\ \min_{{j\in a^{nf}-i}}|h_{j\vers a}|\quad\textrm{if }a\vers i\textrm{ floppy}
\end{split}
\end{equation}
and the maximum weight $d_1$ is given by:
\begin{equation}
d_1=\lim_{\beta\vers -\infty}
\partial_\beta \phi_\bze(\beta)=\sum_{i\ {\rm  floppy}}^N \frac{1+\mathcal{S}\left(\sum_{a\in i}u_{a\vers i}+1\right)}{2}
\end{equation}

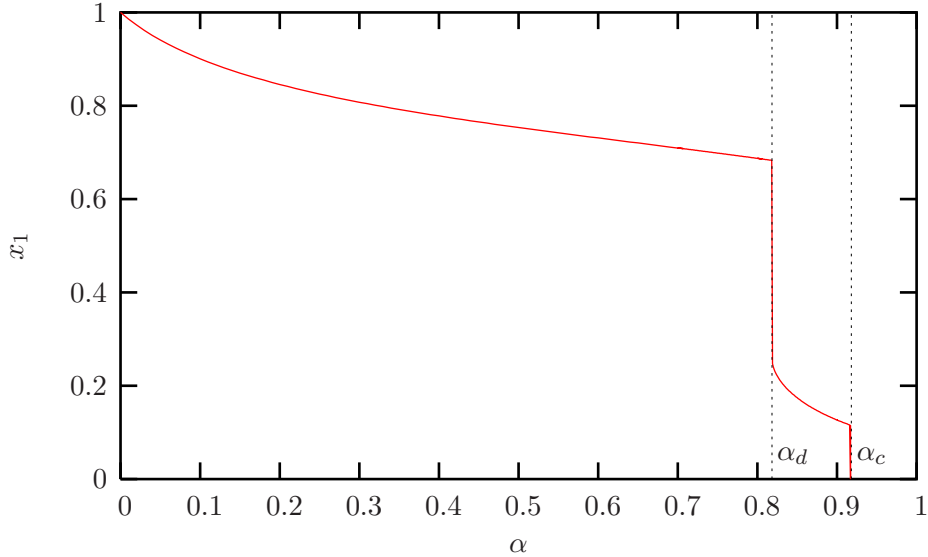
\begin{figure}
\begin{center}
\input{rsmaxdist2.tex}
\caption{\label{fig:rsmaxdist} Diameter of a cluster of solutions. When one decreases $\alpha$ below $\alpha_d$ all clusters aggregate into one big cluster, thus explaining the discontinuity.}
\end{center}
\end{figure}
%ICI Je crois qu'il faut tracer x1 en fonction de alpha, plutot: c'est plus facile a lire. Aussi clarifier la discontinuite des axesou plutot tracer pour tous les alpha.

These equations are presented for single XORSAT formul{\ae}, and can be solved
by simple iteration of the corresponding message-passing rules. In practice
however, in the regime where $\alpha$ is near (but smaller than) $\alpha_d$,
one does not always reach convergence. This is arguably due to the hard nature
of XORSAT constraints, as it was pointed out in \cite{MontanariSemerjian05-2}:
as one nears the dynamical transition, hopping from one solution to the other
requires an increasing (yet sub-extensive) number of changes, making the
sampling of solutions difficult. To circumvent this problem, we can work
directly in the infinite-length limit by considering the probability
distribution functions (pdfs) of each kind of message:
\begin{equation}
\begin{split}
P(h)&=\frac{1}{Mk}\sum_{(i,a)}\delta_{h,h_{i\vers a}}\\
Q(u)&=\frac{1}{Mk}\sum_{(i,a)}\delta_{u,u_{a\vers i}}
\end{split}
\end{equation}
When $N\vers\infty$, self-consistency equations for these distributions read:
\begin{equation}\label{eq:averagediametercav}
\begin{split}
P(h)&=\sum_\ell \pi_{k\alpha w}(\ell) \int \prod_{a=1}^\ell \ud u_a\ Q(u_a)\delta\left(h-\sum_{a=1}^\ell u_a -1\right)\\
Q(u)&=\frac{1}{w}\sum_{i=1}^{k-1} \binom{k-1}{i}v^i (1-v)^{k-1-i} \int \prod_{j=1}^i\ud h_j\ P(h_j)\delta\left[u+\mathcal{S}\left(\prod_{j=1}^i (- h_{j})\right)\min_{j}|h_{j}|\right]
\end{split}
\end{equation}
and one has:
\begin{equation}
x_1(\alpha)=\lim_{N\vers \infty}\frac{d_1}{N}=\e^{-\lambda}\int \ud h\,P(h)\ \frac{1+\mathcal{S}(h)}{2}
\end{equation}
These equations can be solved with a population dynamics algorithm \cite{MezardParisi01}. In Fig.~\eqref{fig:rsmaxdist}, we represent the maximal diameter $x_1$ as a function of $\alpha$.

\section{Minimal and maximal distances between clusters}\label{sec:inter}

In section \ref{sec:single} we have set up the formalism for computing the minimal and the maximal weights in a given cluster $\bc$ using the cavity method. In order to evaluate the minimal and maximal weights in {\em all} clusters expect $\bze$, we resort to a statistical treatment of the cavity equations. This scheme is known as the 1RSB cavity method in the replica language.
We first specialize to the case of minimal weights, the other case being formally equivalent. We already know that the number of clusters grows exponentially with $N$. Here we further assume that the number of clusters with a given minimal weight $x_\bc$ is exponential in $N$, and we define the complexity
\begin{equation}
\sum_{\bc\neq \bze}\delta(x,x_\bc)=2^{N\Sigma_m(x)}.
\end{equation}
To this quantity we associate the 1RSB potential
\begin{equation}
2^{N\psi_m(y)}=\sum_{\bc\neq\bze}2^{-Ny x_\bc}=\int \ud x\ 2^{N(\Sigma_m(x)-yx)}.
\end{equation}
When $N$ is large, a saddle-point evaluation of this quantity yields:
\begin{equation}
\psi_m(y)=\min_x \left[yx-\Sigma_m(x)\right]=yx^*-\Sigma_m(x^*)\quad\textrm{with}\quad y=\partial_x \Sigma_m(x^*)
\end{equation}
$\psi_m(y)$ is thus related to $\Sigma_m(x)$ by a Legendre transformation. In terms of statistical mechanics, $m$ is an inverse temperature coupled to the ``energy'' $x_\bc$; the complexity plays the role of a micro-canonical entropy, and the potential is equivalent to a free energy, up to a factor $m$.
The minimal weight in all clusters (expect $\bze$)
is given by the smallest $x$ such that $\Sigma_m(x)\geq 0$. Our goal is now to compute $\psi_m(y)$, and to infer $\Sigma_m(x)$ by inverse Legendre transformation.

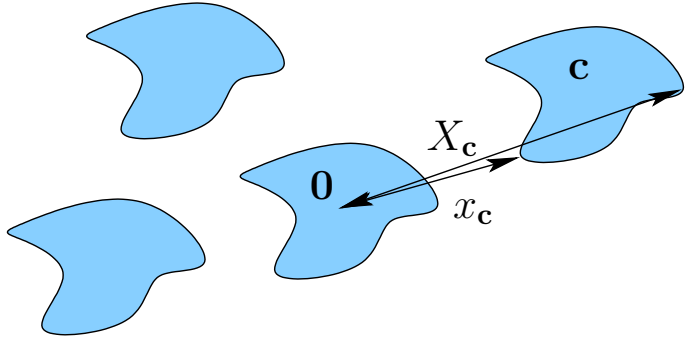
\begin{figure}
\begin{center}
\resizebox{.55\linewidth}{!}{\input{clusterdistance.pstex_t}}
\caption{\label{fig:cldist} Pictorial representation of the clustered space of solutions around $\bze$ in the $N$-dimensional hypercube. For a cluster $\bc$, the minimal and maximal distances $x_\bc$ and $X_\bc$ are depicted.
}
\end{center}
\end{figure}

We proceed to the statistical analysis of the cavity equations under Boltzmann measure $2^{-Nyx_\bc}$. This amounts to writing 1RSB cavity equations, where messages are distributions of RS messages over all clusters.
The distribution of messages on floppy edges is described by the two pdfs:
\begin{eqnarray}
P^{i\vers a}(h)&=&\left\langle \delta(h,h^\bc_{i\vers a})\right\rangle\\
Q^{a\vers i}(u)&=&\left\langle\delta(u,u^\bc_{a\vers i})\right\rangle.
\end{eqnarray}
The average $\langle \cdot \rangle$ is performed with the aforementioned measure on clusters, with the implicit assumption that the edge $(i,a)$ has been removed.
On frozen edges, messages are trivial, but their values depend on the considered cluster. We thus define for frozen edges:
\begin{eqnarray}
P^{i\vers a}_{0}&=&\left\langle \delta(p^0_{i\vers a},1)\right\rangle\qquad P^{i\vers a}_{1}=1-P^{i\vers a}_{0}\\
Q^{a\vers i}_{0}&=&\left\langle \delta(q^0_{a\vers i},1)\right\rangle\qquad Q^{a\vers i}_{1}=1-Q^{a\vers i}_{0}
\end{eqnarray}

In order to write a closed set of equations for these probability distributions, we need to know how the Boltzmann weight $2^{-Nyx_\bc}$ biases the message-passing procedure: when a field $h_{i\vers a}$ is estimated as a function of its ``grand-parents'' ($\{h_{j\vers b}\},\ j\in b-i,\ b\in i-a$), a re-weighting term $2^{-y\Delta x_{i\vers a}}$ is associated to it \cite{MezardParisi01,MezardZecchina02}, where $\Delta x_{i\vers a}$ is the contribution of $i$ and its adjacent checks (except $a$) to the total weight. 
This contribution is obtained as $\Delta x_{i+a\in i}$ in Eq.~\eqref{eq:potcontrib1}-\eqref{eq:potcontrib3}, but with $a$ removed.

The 1RSB cavity equations read:
\begin{itemize}
\item $i\vers a$ frozen:
\begin{equation}
\begin{split}
P^{i\vers a}_{0}=&\frac{1}{\mathcal{Z}_{i\vers a}}\prod_{b\in i^f-a}Q^{b\vers i}_{0}\int \prod_{b\in i^{nf}-a}\ud u_{b\vers i} Q^{b\vers i}(u_{b\vers i})2^{-y\sum_{b\in i^{nf}-a} |u_{b\vers i}|\vartheta(-u_{b\vers i})}\\
P^{i\vers a}_{1}=&\frac{1}{\mathcal{Z}_{i\vers a}}\prod_{b\in i^f-a}Q^{b\vers i}_{1}\int \prod_{b\in i^{nf}-a}\ud u_{b\vers i} Q^{b\vers i}(u_{b\vers i})2^{-y\left(1+\sum_{b\in i^{nf}-a} |u_{b\vers i}|\vartheta(u_{b\vers i})\right)}
\end{split}
\end{equation}
\item $i\vers a$ floppy:
\begin{equation}
\begin{split}
P^{i\vers a}(h)=&\frac{1}{\mathcal{Z}_{i\vers a}}\int \prod_{b\in i-a}\ud u_{b\vers i} Q^{b\vers i}(u_{b\vers i})2^{-y/2\left(\sum_{b\in i-a}|u_{b\vers i}|+1-|\sum_{b\in i-a}u_{b\vers i}+1|\right)}\\ &\times \delta\left(h-1-\sum_{b\in i-a}u_{b\vers i}\right)
\end{split}
\end{equation}
(here and in the previous equations $\mathcal{Z}_{i\vers a}$ is a normalization constant)
\item $a\vers i$ frozen:
\begin{equation}
Q^{a\vers i}_{0}=\frac{1+\prod_{j\in a-i}(2P^{j\vers a}_{0}-1)}{2}\\
\end{equation}
\item $a\vers i$ floppy:
\begin{equation}
\begin{split}
Q^{a\vers i}(u)=&\sum_{\substack{\{c_j=0,1\}\\j\in a^f-i}}\prod_{j\in a^f-i}P^{j\vers a}_{c_j}\int \prod_{j\in a^{nf}-i}\ud h_{j\vers a}P^{j\vers a}(h_{j\vers a})\\ &\times \delta\left[u-\mathcal{S}\left(\prod_{j\in a^{nf}-i} h_{j\vers a}\prod_{j\in a^{f}-i}(-1)^{c_j}\right)\ \min_{{j\in a^{nf}-i}}|h_{j\vers a}|\right]
\end{split}
\end{equation}
\end{itemize}
The potential $\psi_m(y)$ is obtained by a Bethe-like formula \cite{MezardZecchina02}:
\begin{equation}
N\psi_m(y)=\sum_i \Delta \psi_{i+a\in i}-(k-1)\sum_{a} \Delta \psi_a
\end{equation}
with
\begin{equation}\label{eq:phiintermin1}
\begin{split}
\Delta \psi_{i+a\in i}=&-\log \left\langle 2^{-y\Delta x_{i+a\in i}}\right\rangle =-\log \mathcal{Z}_{i+a\in i}\\
\Delta \psi_a=&-\log \left\langle 2^{-y\Delta x_a}\right\rangle\\
=&-\log \frac{1+\prod_{i\in a}(2P^{i\vers a}_{0}-1)}{2}\quad\textrm{if }a\in\textrm{core}\\
=&-\log\sum_{\substack{\{c_i=0,1\}\\i\in a^f}}\prod_{j\in a^f}P^{i\vers a}_{c_i}\int \prod_{i\in a^{nf}}\ud h_{i\vers a}P^{i\vers a}(h_{i\vers a})\\&\times \exp{\left[-y\log(2)\vartheta\left(-\prod_{i\in a^{nf}} h_{i\vers a}\prod_{i\in a^{f}}(-1)^{c_i}\right)\ \min_{{i\in a^{nf}}}|h_{i\vers a}|\right]}\quad\textrm{otherwise}
\end{split}
\end{equation}
where $\mathcal{Z}_{i+a\in i}$ is defined as $\mathcal{Z}_{i\vers a}$ but in the presence of $a$.

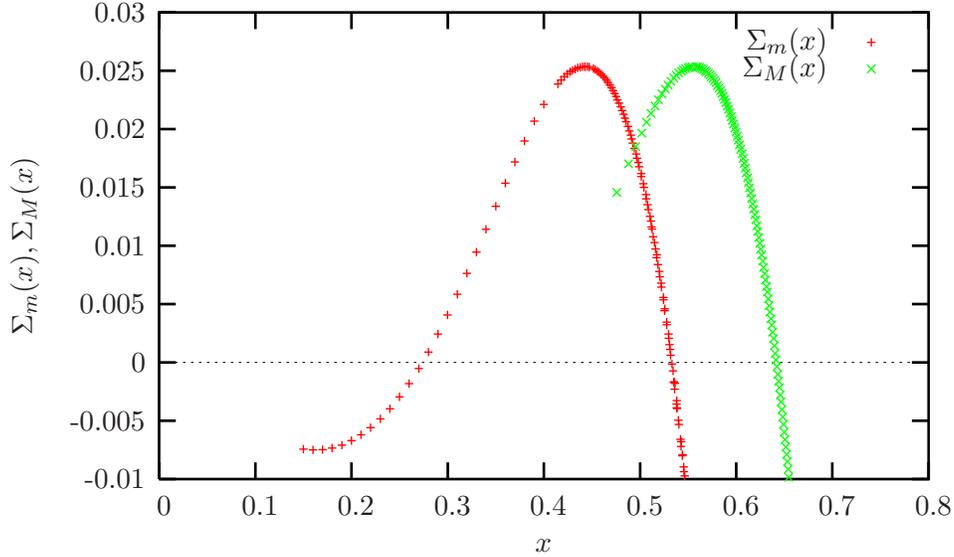
\begin{figure}
\begin{center}
\input{wefk3.tex}
\caption{\label{fig:wefk3} Minimal and maximal distance complexities as a function of the reduced distance $x$, for $k=3$, $N=10000$ and $M=8600$.}
\end{center}
\end{figure}

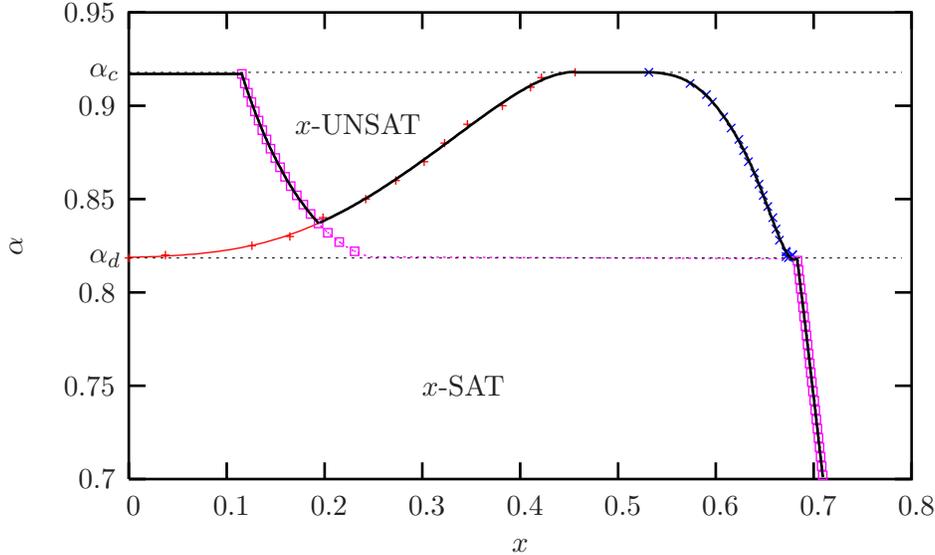
\begin{figure}
\begin{center}
\input{alphaxphasek3.tex}
\caption{\label{fig:alphaxphasek3} Phase diagram of the $3$-XORSAT problem in the $(x,\alpha)$ plane. The cluster diameter ($\square$), as well as minimal ($+$) and maximal ($\times$) distances between solutions of distinct clusters, are represented. The thick line is the $x$-satisfiability threshold.
}
\end{center}
\end{figure}

Like in the diameter calculation, 1RSB cavity equations can be interpreted as message-passing update rules, with the difference that messages are now surveys over all clusters. The output of that procedure is the minimal distance complexity $\Sigma_m(x)$, obtained as the inverse Legendre transform of $\psi_m(y)$.
We refer to the corresponding algorithm as ``distance survey propagation''. The same procedure can be implemented in the $\beta\vers -\infty$ limit, and yields the maximal distance complexity:
\begin{equation}
\Sigma_M(x)=\frac{1}{N}\log \sum_{\bc\neq \bze}\delta(x,X_\bc),
\end{equation}
where $X_\bc$ is the maximal weight in cluster $\bc$ (see Fig.~\ref{fig:cldist}).
Note that in the particular case where $y=0$, which corresponds to a uniform measure over the clusters, classical SP is recovered for both versions of the algorithm (minimal and maximal distance): in that limit we have $Q^{a\vers i}_0=P^{i\vers a}_0=1/2$, and the calculation of $\psi_m(0)$ and $\psi_M(0)$ gives back $-\Sigma(\alpha)$, the total complexity \eqref{eq:complexity}, as expected.

The practical implementation of distance-SP demands particular care when small
distances are considered: it turns out that distance complexities
$\Sigma_m(x)$ and $\Sigma_M(x)$ are not concave, which entails that the
functions $\psi_m(y)$ and $\psi_M(y)$ are multivalued in a certain range of
$y$. A way to circumvent this problem (already used in \cite{DiMontanari04})
is to keep the weight $x=\partial_y \psi_m(y)$ fixed after each iteration, and
to deduce $y$ accordingly. Here is how the algorithm proceeds for a given
reduced weight $x$:%ICI
\begin{enumerate}
\item Run classical SP.
\item Initialize all floppy and frozen messages $\{P_{i\vers a}\}$, $\{Q_{a\vers i}\}$ to random values. Choose a (reasonable) value for $y$.
\item Until convergence is reached, do:
\begin{itemize}
\item Update all $a\vers i$ messages $\{Q_{a\vers i}\}$, and then all $i\vers a$ messages $\{P_{i\vers a}\}$ at inverse temperature $y$.
\item Find $y$ such that $x=\partial_y\psi_m(y,\{P_{i\vers a}\},\{Q_{a\vers i}\})$ by the secant method, $\{P_{i\vers a}\}$ and $\{Q_{a\vers i}\}$ being fixed.
\end{itemize}
\item Compute $\psi_m(y,\{P_{i\vers a}\},\{Q_{a\vers i}\})$ as well as its derivative and deduce $\Sigma_m(x)=yx-\psi_m(y)$.
\end{enumerate}
Note that since the messages are pdfs themselves, the update of each of them in step 3 is performed by a population dynamics sub-routine.

Fig.~\ref{fig:wefk3} shows the minimal and maximal weight complexities $\Sigma_m(x)$ and $\Sigma_M(x)$ for a random $3$-XORSAT formula with $N=10000$ and $M=8600$. These complexities can be regarded as kinds of weight enumerator functions for clusters. Their fluctuations from formula to formula can be significant (15\%), even for large system sizes ($N=10000$).

An average version (density evolution) of distance-SP can also be implemented for random $k$-XORSAT, in the same spirit as Eq.~\eqref{eq:averagediametercav}. Such a computation involves distributions (on edges) of distributions (on clusters), and can be solved by population dynamics, where each element of the population is itself a population. The zeros of $\overline{\Sigma_m(x)}$ and $\overline{\Sigma_M(x)}$ thus obtained yield the minimal and maximal inter-cluster distances $x_2(\alpha)$ and $x_3(\alpha)$, respectively, as shown in Fig.~\ref{fig:alphaxphasek3}. Together with the cluster diameter $x_1(\alpha)$ computed in section \ref{sec:diameter}, these values are used to construct the $x$-satisfiability threshold.

Our algorithm can in principle be run on any system of Boolean
linear equations, and is expected to give reasonable results provided
that the loops of the underlying Tanner graph are large.
The case of LDPC codes is of particular interest because it allows
several simplifications and has been extensively studied from both the
combinatorial \cite{DiProietti02} and statistical physics
\cite{Condamin02,DiMontanari04} point of view. LDPC codes are homogeneous Boolean
linear systems where parity checks and
variables may have arbitrary degree distributions, with the
restriction that variables should always have degrees no less
than $2$. This implies that the leaf removal
algorithm is inefficient on such linear systems: all variables belong
to the core, and are frozen. In particular, each cluster is made of
one unique solution: the cluster diameter is $0$, and the minimal and
maximal inter-cluster distances coincide. Their common
complexity $\Sigma_m(x)=\Sigma_M(x)$ is often called  `weight
enumerator exponent' and is an important property of ensembles of codes. 
Translated into our formalism, this means that all messages are frozen,
and the distance-SP
algorithm simplifies dramatically:
\begin{equation}
P^{i\vers a}_{0}=\frac{1}{\mathcal{Z}_{i\vers a}}\prod_{b\in
  i^f-a}Q^{b\vers i}_{0},\qquad
P^{i\vers a}_{1}=\frac{1}{\mathcal{Z}_{i\vers a}}\prod_{b\in
  i^f-a}Q^{b\vers i}_{1}
2^{-y}
\end{equation}
\begin{equation}
Q^{a\vers i}_{0}=\frac{1+\prod_{j\in a-i}(2P^{j\vers a}_{0}-1)}{2}\\
\end{equation}

Not surprisingly, the density
evolution analysis of this simplified algorithm yields the same equations as those obtained with the replica
method in \cite{Condamin02,DiMontanari04}.
 
\section{Conclusion and discussion}\label{sec:conclu}
We have applied the cavity method to estimate extremal distances
between solutions of random linear systems with large girth in the
clustered phase. Our results
are used to compute the $x$-satisfiability threshold of the
random $k$-XORSAT problem. The notion of $x$-satisfiability, which
tells us whether one can find a pair of solutions separated by a
Hamming distance $x$, was introduced in the
context of another constraint satisfaction problem, $k$-SAT, where it
was used to give rigorous evidence in favor of the clustering
phenomenon \cite{MezardMora05}.

Although $k$-XORSAT is a rather simple problem, it
displays a very similar phase diagram as harder problems such as
$k$-SAT or $q$-colorability. In particular, its clustered phase is
well defined and understood.
That said, finding extremal distances in the solution space of linear Boolean equations
is a hard task in general: for instance, the decision problem associated with finding the minimal weight of LDPC codes is NP-complete
\cite{Vardy97}.

We were able to compute three quantities: the cluster
diameter, as well as the minimal and maximal inter-cluster distances. We believe our method to give a good approximation for systems with large girth, and to be exact in the thermodynamic limit for random XORSAT.
In
the line of Survey Propagation,
we devised
a series of algorithms for these tasks, which explicitly exploit the
clustered structure of the solution space. More precisely, 
the space of solutions is characterized by two hierarchical levels of
fluctuations: inside and between clusters. In $k$-XORSAT, these two kinds
of fluctuations are carried by two disjoint sets of variables, and
our algorithms
explicitly distinguish between these two types of variables.
In the special case of LDPC codes, the point-like nature of
clusters much simplifies the equations, and previous
expressions of the weight enumerator exponent obtained by the replica
method are recovered.

The method presented here offers a number of generalizations. In
particular, it could be used at finite temperature to yield
the full weight enumerator function. More interestingly,
it could be adapted to deal with other CSN, such as $k$-SAT, for which only bounds are known;
unfortunately, numerical computations are in that case much heavier, albeit formally similar.
Let us mention that a similar approach
was followed in \cite{MezardPalassini05} in the case of
$q$-colorability, with the difference that distances were
estimated from a reference configuration (which is not a solution)
instead of considering distances between solutions.

Our work studies the geometrical properties
of the solution space by taking explicitly into account
fluctuations inside clusters, captured by the `evanescent
fields'. This very general approach, already explored in \cite{MezardPalassini05}, allows to gain a better
understanding of the fine structure of the clustered phase, and seems to us a promising direction for future work. Also, with similar
tools, decimation schemes such as the one introduced in
\cite{MezardZecchina02} could be used to select solutions or clusters with
particular properties.

\bigskip

We would like to thank Andrea Montanari for sharing the
numerical trick used in the replica evaluation of the weight
enumerator function of LDPC codes \cite{DiMontanari04}.
This work has been supported in part by the EU through the network
MTR 2002-00319 `STIPCO' and the FP6 IST consortium `EVERGROW'.

\bibliographystyle{unsrt}

\bibliography{myentries}

\end{document}

%% file: rsmaxdist2.tex
% GNUPLOT: LaTeX picture with Postscript
\begingroup%
  \makeatletter%
  \newcommand{\GNUPLOTspecial}{%
    \@sanitize\catcode`\%=14\relax\special}%
  \setlength{\unitlength}{0.1bp}%
{\GNUPLOTspecial{!
%!PS-Adobe-2.0
%%Title: rsmaxdist2.tex
%%Creator: gnuplot 3.7 patchlevel 3
%%CreationDate: Mon Sep  4 10:24:31 2006
%%DocumentFonts: 
%%BoundingBox: 0 0 360 216
%%Orientation: Landscape
%%Pages: (atend)
%%EndComments
/gnudict 256 dict def
gnudict begin
/Color true def
/Solid false def
/gnulinewidth 5.000 def
/userlinewidth gnulinewidth def
/vshift -33 def
/dl {10 mul} def
/hpt_ 31.5 def
/vpt_ 31.5 def
/hpt hpt_ def
/vpt vpt_ def
/M {moveto} bind def
/L {lineto} bind def
/R {rmoveto} bind def
/V {rlineto} bind def
/vpt2 vpt 2 mul def
/hpt2 hpt 2 mul def
/Lshow { currentpoint stroke M
  0 vshift R show } def
/Rshow { currentpoint stroke M
  dup stringwidth pop neg vshift R show } def
/Cshow { currentpoint stroke M
  dup stringwidth pop -2 div vshift R show } def
/UP { dup vpt_ mul /vpt exch def hpt_ mul /hpt exch def
  /hpt2 hpt 2 mul def /vpt2 vpt 2 mul def } def
/DL { Color {setrgbcolor Solid {pop []} if 0 setdash }
 {pop pop pop Solid {pop []} if 0 setdash} ifelse } def
/BL { stroke userlinewidth 2 mul setlinewidth } def
/AL { stroke userlinewidth 2 div setlinewidth } def
/UL { dup gnulinewidth mul /userlinewidth exch def
      dup 1 lt {pop 1} if 10 mul /udl exch def } def
/PL { stroke userlinewidth setlinewidth } def
/LTb { BL [] 0 0 0 DL } def
/LTa { AL [1 udl mul 2 udl mul] 0 setdash 0 0 0 setrgbcolor } def
/LT0 { PL [] 1 0 0 DL } def
/LT1 { PL [4 dl 2 dl] 0 1 0 DL } def
/LT2 { PL [2 dl 3 dl] 0 0 1 DL } def
/LT3 { PL [1 dl 1.5 dl] 1 0 1 DL } def
/LT4 { PL [5 dl 2 dl 1 dl 2 dl] 0 1 1 DL } def
/LT5 { PL [4 dl 3 dl 1 dl 3 dl] 1 1 0 DL } def
/LT6 { PL [2 dl 2 dl 2 dl 4 dl] 0 0 0 DL } def
/LT7 { PL [2 dl 2 dl 2 dl 2 dl 2 dl 4 dl] 1 0.3 0 DL } def
/LT8 { PL [2 dl 2 dl 2 dl 2 dl 2 dl 2 dl 2 dl 4 dl] 0.5 0.5 0.5 DL } def
/Pnt { stroke [] 0 setdash
   gsave 1 setlinecap M 0 0 V stroke grestore } def
/Dia { stroke [] 0 setdash 2 copy vpt add M
  hpt neg vpt neg V hpt vpt neg V
  hpt vpt V hpt neg vpt V closepath stroke
  Pnt } def
/Pls { stroke [] 0 setdash vpt sub M 0 vpt2 V
  currentpoint stroke M
  hpt neg vpt neg R hpt2 0 V stroke
  } def
/Box { stroke [] 0 setdash 2 copy exch hpt sub exch vpt add M
  0 vpt2 neg V hpt2 0 V 0 vpt2 V
  hpt2 neg 0 V closepath stroke
  Pnt } def
/Crs { stroke [] 0 setdash exch hpt sub exch vpt add M
  hpt2 vpt2 neg V currentpoint stroke M
  hpt2 neg 0 R hpt2 vpt2 V stroke } def
/TriU { stroke [] 0 setdash 2 copy vpt 1.12 mul add M
  hpt neg vpt -1.62 mul V
  hpt 2 mul 0 V
  hpt neg vpt 1.62 mul V closepath stroke
  Pnt  } def
/Star { 2 copy Pls Crs } def
/BoxF { stroke [] 0 setdash exch hpt sub exch vpt add M
  0 vpt2 neg V  hpt2 0 V  0 vpt2 V
  hpt2 neg 0 V  closepath fill } def
/TriUF { stroke [] 0 setdash vpt 1.12 mul add M
  hpt neg vpt -1.62 mul V
  hpt 2 mul 0 V
  hpt neg vpt 1.62 mul V closepath fill } def
/TriD { stroke [] 0 setdash 2 copy vpt 1.12 mul sub M
  hpt neg vpt 1.62 mul V
  hpt 2 mul 0 V
  hpt neg vpt -1.62 mul V closepath stroke
  Pnt  } def
/TriDF { stroke [] 0 setdash vpt 1.12 mul sub M
  hpt neg vpt 1.62 mul V
  hpt 2 mul 0 V
  hpt neg vpt -1.62 mul V closepath fill} def
/DiaF { stroke [] 0 setdash vpt add M
  hpt neg vpt neg V hpt vpt neg V
  hpt vpt V hpt neg vpt V closepath fill } def
/Pent { stroke [] 0 setdash 2 copy gsave
  translate 0 hpt M 4 {72 rotate 0 hpt L} repeat
  closepath stroke grestore Pnt } def
/PentF { stroke [] 0 setdash gsave
  translate 0 hpt M 4 {72 rotate 0 hpt L} repeat
  closepath fill grestore } def
/Circle { stroke [] 0 setdash 2 copy
  hpt 0 360 arc stroke Pnt } def
/CircleF { stroke [] 0 setdash hpt 0 360 arc fill } def
/C0 { BL [] 0 setdash 2 copy moveto vpt 90 450  arc } bind def
/C1 { BL [] 0 setdash 2 copy        moveto
       2 copy  vpt 0 90 arc closepath fill
               vpt 0 360 arc closepath } bind def
/C2 { BL [] 0 setdash 2 copy moveto
       2 copy  vpt 90 180 arc closepath fill
               vpt 0 360 arc closepath } bind def
/C3 { BL [] 0 setdash 2 copy moveto
       2 copy  vpt 0 180 arc closepath fill
               vpt 0 360 arc closepath } bind def
/C4 { BL [] 0 setdash 2 copy moveto
       2 copy  vpt 180 270 arc closepath fill
               vpt 0 360 arc closepath } bind def
/C5 { BL [] 0 setdash 2 copy moveto
       2 copy  vpt 0 90 arc
       2 copy moveto
       2 copy  vpt 180 270 arc closepath fill
               vpt 0 360 arc } bind def
/C6 { BL [] 0 setdash 2 copy moveto
      2 copy  vpt 90 270 arc closepath fill
              vpt 0 360 arc closepath } bind def
/C7 { BL [] 0 setdash 2 copy moveto
      2 copy  vpt 0 270 arc closepath fill
              vpt 0 360 arc closepath } bind def
/C8 { BL [] 0 setdash 2 copy moveto
      2 copy vpt 270 360 arc closepath fill
              vpt 0 360 arc closepath } bind def
/C9 { BL [] 0 setdash 2 copy moveto
      2 copy  vpt 270 450 arc closepath fill
              vpt 0 360 arc closepath } bind def
/C10 { BL [] 0 setdash 2 copy 2 copy moveto vpt 270 360 arc closepath fill
       2 copy moveto
       2 copy vpt 90 180 arc closepath fill
               vpt 0 360 arc closepath } bind def
/C11 { BL [] 0 setdash 2 copy moveto
       2 copy  vpt 0 180 arc closepath fill
       2 copy moveto
       2 copy  vpt 270 360 arc closepath fill
               vpt 0 360 arc closepath } bind def
/C12 { BL [] 0 setdash 2 copy moveto
       2 copy  vpt 180 360 arc closepath fill
               vpt 0 360 arc closepath } bind def
/C13 { BL [] 0 setdash  2 copy moveto
       2 copy  vpt 0 90 arc closepath fill
       2 copy moveto
       2 copy  vpt 180 360 arc closepath fill
               vpt 0 360 arc closepath } bind def
/C14 { BL [] 0 setdash 2 copy moveto
       2 copy  vpt 90 360 arc closepath fill
               vpt 0 360 arc } bind def
/C15 { BL [] 0 setdash 2 copy vpt 0 360 arc closepath fill
               vpt 0 360 arc closepath } bind def
/Rec   { newpath 4 2 roll moveto 1 index 0 rlineto 0 exch rlineto
       neg 0 rlineto closepath } bind def
/Square { dup Rec } bind def
/Bsquare { vpt sub exch vpt sub exch vpt2 Square } bind def
/S0 { BL [] 0 setdash 2 copy moveto 0 vpt rlineto BL Bsquare } bind def
/S1 { BL [] 0 setdash 2 copy vpt Square fill Bsquare } bind def
/S2 { BL [] 0 setdash 2 copy exch vpt sub exch vpt Square fill Bsquare } bind def
/S3 { BL [] 0 setdash 2 copy exch vpt sub exch vpt2 vpt Rec fill Bsquare } bind def
/S4 { BL [] 0 setdash 2 copy exch vpt sub exch vpt sub vpt Square fill Bsquare } bind def
/S5 { BL [] 0 setdash 2 copy 2 copy vpt Square fill
       exch vpt sub exch vpt sub vpt Square fill Bsquare } bind def
/S6 { BL [] 0 setdash 2 copy exch vpt sub exch vpt sub vpt vpt2 Rec fill Bsquare } bind def
/S7 { BL [] 0 setdash 2 copy exch vpt sub exch vpt sub vpt vpt2 Rec fill
       2 copy vpt Square fill
       Bsquare } bind def
/S8 { BL [] 0 setdash 2 copy vpt sub vpt Square fill Bsquare } bind def
/S9 { BL [] 0 setdash 2 copy vpt sub vpt vpt2 Rec fill Bsquare } bind def
/S10 { BL [] 0 setdash 2 copy vpt sub vpt Square fill 2 copy exch vpt sub exch vpt Square fill
       Bsquare } bind def
/S11 { BL [] 0 setdash 2 copy vpt sub vpt Square fill 2 copy exch vpt sub exch vpt2 vpt Rec fill
       Bsquare } bind def
/S12 { BL [] 0 setdash 2 copy exch vpt sub exch vpt sub vpt2 vpt Rec fill Bsquare } bind def
/S13 { BL [] 0 setdash 2 copy exch vpt sub exch vpt sub vpt2 vpt Rec fill
       2 copy vpt Square fill Bsquare } bind def
/S14 { BL [] 0 setdash 2 copy exch vpt sub exch vpt sub vpt2 vpt Rec fill
       2 copy exch vpt sub exch vpt Square fill Bsquare } bind def
/S15 { BL [] 0 setdash 2 copy Bsquare fill Bsquare } bind def
/D0 { gsave translate 45 rotate 0 0 S0 stroke grestore } bind def
/D1 { gsave translate 45 rotate 0 0 S1 stroke grestore } bind def
/D2 { gsave translate 45 rotate 0 0 S2 stroke grestore } bind def
/D3 { gsave translate 45 rotate 0 0 S3 stroke grestore } bind def
/D4 { gsave translate 45 rotate 0 0 S4 stroke grestore } bind def
/D5 { gsave translate 45 rotate 0 0 S5 stroke grestore } bind def
/D6 { gsave translate 45 rotate 0 0 S6 stroke grestore } bind def
/D7 { gsave translate 45 rotate 0 0 S7 stroke grestore } bind def
/D8 { gsave translate 45 rotate 0 0 S8 stroke grestore } bind def
/D9 { gsave translate 45 rotate 0 0 S9 stroke grestore } bind def
/D10 { gsave translate 45 rotate 0 0 S10 stroke grestore } bind def
/D11 { gsave translate 45 rotate 0 0 S11 stroke grestore } bind def
/D12 { gsave translate 45 rotate 0 0 S12 stroke grestore } bind def
/D13 { gsave translate 45 rotate 0 0 S13 stroke grestore } bind def
/D14 { gsave translate 45 rotate 0 0 S14 stroke grestore } bind def
/D15 { gsave translate 45 rotate 0 0 S15 stroke grestore } bind def
/DiaE { stroke [] 0 setdash vpt add M
  hpt neg vpt neg V hpt vpt neg V
  hpt vpt V hpt neg vpt V closepath stroke } def
/BoxE { stroke [] 0 setdash exch hpt sub exch vpt add M
  0 vpt2 neg V hpt2 0 V 0 vpt2 V
  hpt2 neg 0 V closepath stroke } def
/TriUE { stroke [] 0 setdash vpt 1.12 mul add M
  hpt neg vpt -1.62 mul V
  hpt 2 mul 0 V
  hpt neg vpt 1.62 mul V closepath stroke } def
/TriDE { stroke [] 0 setdash vpt 1.12 mul sub M
  hpt neg vpt 1.62 mul V
  hpt 2 mul 0 V
  hpt neg vpt -1.62 mul V closepath stroke } def
/PentE { stroke [] 0 setdash gsave
  translate 0 hpt M 4 {72 rotate 0 hpt L} repeat
  closepath stroke grestore } def
/CircE { stroke [] 0 setdash 
  hpt 0 360 arc stroke } def
/Opaque { gsave closepath 1 setgray fill grestore 0 setgray closepath } def
/DiaW { stroke [] 0 setdash vpt add M
  hpt neg vpt neg V hpt vpt neg V
  hpt vpt V hpt neg vpt V Opaque stroke } def
/BoxW { stroke [] 0 setdash exch hpt sub exch vpt add M
  0 vpt2 neg V hpt2 0 V 0 vpt2 V
  hpt2 neg 0 V Opaque stroke } def
/TriUW { stroke [] 0 setdash vpt 1.12 mul add M
  hpt neg vpt -1.62 mul V
  hpt 2 mul 0 V
  hpt neg vpt 1.62 mul V Opaque stroke } def
/TriDW { stroke [] 0 setdash vpt 1.12 mul sub M
  hpt neg vpt 1.62 mul V
  hpt 2 mul 0 V
  hpt neg vpt -1.62 mul V Opaque stroke } def
/PentW { stroke [] 0 setdash gsave
  translate 0 hpt M 4 {72 rotate 0 hpt L} repeat
  Opaque stroke grestore } def
/CircW { stroke [] 0 setdash 
  hpt 0 360 arc Opaque stroke } def
/BoxFill { gsave Rec 1 setgray fill grestore } def
/Symbol-Oblique /Symbol findfont [1 0 .167 1 0 0] makefont
dup length dict begin {1 index /FID eq {pop pop} {def} ifelse} forall
currentdict end definefont pop
end
}}%
\begin{picture}(3600,2160)(0,0)%
{\GNUPLOTspecial{"
gnudict begin
gsave
0 0 translate
0.100 0.100 scale
0 setgray
newpath
1.000 UL
LTb
450 300 M
63 0 V
2937 0 R
-63 0 V
450 652 M
63 0 V
2937 0 R
-63 0 V
450 1004 M
63 0 V
2937 0 R
-63 0 V
450 1356 M
63 0 V
2937 0 R
-63 0 V
450 1708 M
63 0 V
2937 0 R
-63 0 V
450 2060 M
63 0 V
2937 0 R
-63 0 V
450 300 M
0 63 V
0 1697 R
0 -63 V
750 300 M
0 63 V
0 1697 R
0 -63 V
1050 300 M
0 63 V
0 1697 R
0 -63 V
1350 300 M
0 63 V
0 1697 R
0 -63 V
1650 300 M
0 63 V
0 1697 R
0 -63 V
1950 300 M
0 63 V
0 1697 R
0 -63 V
2250 300 M
0 63 V
0 1697 R
0 -63 V
2550 300 M
0 63 V
0 1697 R
0 -63 V
2850 300 M
0 63 V
0 1697 R
0 -63 V
3150 300 M
0 63 V
0 1697 R
0 -63 V
3450 300 M
0 63 V
0 1697 R
0 -63 V
1.000 UL
LTb
450 300 M
3000 0 V
0 1760 V
-3000 0 V
450 300 L
1.000 UL
LT0
450 2060 M
30 -25 V
30 -22 V
30 -21 V
30 -19 V
30 -17 V
30 -16 V
30 -15 V
30 -14 V
30 -13 V
30 -13 V
30 -12 V
30 -11 V
30 -11 V
30 -10 V
30 -10 V
30 -9 V
30 -9 V
30 -8 V
30 -9 V
30 -8 V
30 -7 V
30 -8 V
30 -7 V
30 -7 V
30 -7 V
30 -6 V
30 -7 V
30 -6 V
30 -6 V
30 -6 V
30 -5 V
30 -6 V
30 -5 V
30 -6 V
30 -5 V
30 -5 V
30 -5 V
30 -5 V
30 -5 V
30 -4 V
30 -5 V
30 -5 V
30 -4 V
30 -5 V
30 -4 V
30 -4 V
30 -5 V
30 -4 V
30 -4 V
30 -4 V
30 -4 V
30 -4 V
30 -4 V
30 -4 V
30 -4 V
30 -4 V
30 -4 V
30 -4 V
30 -4 V
30 -3 V
30 -4 V
30 -4 V
30 -4 V
30 -4 V
30 -3 V
30 -4 V
30 -4 V
30 -4 V
30 -4 V
33 -4 V
3 0 V
3 0 V
3 -1 V
3 0 V
3 0 V
3 -1 V
-21 3 V
24 -3 V
3 -1 V
3 0 V
3 0 V
3 -1 V
3 0 V
3 -1 V
3 0 V
3 0 V
3 -1 V
3 0 V
3 0 V
3 -1 V
3 0 V
3 -1 V
3 0 V
3 0 V
3 -1 V
3 0 V
3 -1 V
3 0 V
3 0 V
3 -1 V
3 0 V
3 -1 V
3 0 V
3 0 V
3 -1 V
3 0 V
3 0 V
3 -1 V
3 0 V
3 -1 V
3 0 V
3 0 V
3 -1 V
3 0 V
3 -1 V
3 0 V
3 0 V
3 -1 V
3 0 V
3 -1 V
3 0 V
3 0 V
3 -1 V
3 0 V
3 0 V
3 -1 V
3 0 V
3 -1 V
3 0 V
3 0 V
3 -1 V
3 0 V
3 -1 V
3 0 V
3 0 V
3 -1 V
3 0 V
3 -1 V
3 0 V
3 0 V
3 -1 V
3 0 V
3 -1 V
3 0 V
3 0 V
3 -1 V
3 0 V
3 -1 V
3 0 V
3 0 V
3 -1 V
3 0 V
3 -1 V
3 0 V
3 0 V
3 -1 V
3 0 V
3 0 V
3 -1 V
3 0 V
3 -1 V
3 0 V
3 -1 V
3 0 V
3 0 V
3 -1 V
3 0 V
3 -1 V
3 0 V
6 -1 V
3 0 V
3 -1 V
3 0 V
3 0 V
3 -1 V
-18 3 V
21 -3 V
3 -1 V
3 0 V
3 -1 V
3 0 V
3 0 V
3 -1 V
3 0 V
3 -1 V
3 0 V
3 0 V
3 -1 V
3 -767 V
3 -12 V
3 -9 V
3 -8 V
3 -6 V
3 -6 V
3 -5 V
3 -5 V
3 -5 V
3 -5 V
3 -4 V
3 -4 V
3 -4 V
3 -4 V
3 -3 V
3 -4 V
3 -3 V
3 -4 V
3 -3 V
3 -3 V
3 -3 V
3 -3 V
3 -2 V
3 -3 V
3 -3 V
3 -3 V
3 -2 V
3 -2 V
3 -3 V
3 -3 V
3 -2 V
3 -2 V
3 -3 V
3 -2 V
3 -2 V
3 -2 V
3 -3 V
3 -2 V
3 -2 V
3 -2 V
3 -2 V
3 -2 V
3 -2 V
3 -2 V
3 -2 V
3 -2 V
3 -2 V
3 -1 V
3 -2 V
3 -2 V
3 -2 V
3 -1 V
3 -2 V
3 -2 V
3 -1 V
3 -2 V
3 -2 V
3 -1 V
3 -2 V
3 -1 V
3 -2 V
3 -2 V
3 -1 V
3 -2 V
3 -1 V
3 -2 V
3 -1 V
3 -1 V
3 -2 V
3 -1 V
3 -2 V
3 -1 V
3 -1 V
3 -2 V
3 -1 V
3 -1 V
3 -2 V
3 -1 V
3 -1 V
3 -2 V
3 -1 V
6 -2 V
-3 1 V
6 -3 V
3 -1 V
3 -1 V
3 -1 V
3 -1 V
3 -2 V
3 -1 V
3 -1 V
3 -1 V
3 -1 V
3 -1 V
3 -1 V
3 -2 V
3 -1 V
3 -1 V
3 -204 V
0 203 V
1.000 UL
LTa
2905 300 M
0 18 V
0 18 V
0 17 V
0 18 V
0 18 V
0 18 V
0 17 V
0 18 V
0 18 V
0 18 V
0 18 V
0 17 V
0 18 V
0 18 V
0 18 V
0 17 V
0 18 V
0 18 V
0 18 V
0 18 V
0 17 V
0 18 V
0 18 V
0 18 V
0 17 V
0 18 V
0 18 V
0 18 V
0 18 V
0 17 V
0 18 V
0 18 V
0 18 V
0 17 V
0 18 V
0 18 V
0 18 V
0 18 V
0 17 V
0 18 V
0 18 V
0 18 V
0 17 V
0 18 V
0 18 V
0 18 V
0 18 V
0 17 V
0 18 V
0 18 V
0 18 V
0 17 V
0 18 V
0 18 V
0 18 V
0 18 V
0 17 V
0 18 V
0 18 V
0 18 V
0 17 V
0 18 V
0 18 V
0 18 V
0 18 V
0 17 V
0 18 V
0 18 V
0 18 V
0 17 V
0 18 V
0 18 V
0 18 V
0 18 V
0 17 V
0 18 V
0 18 V
0 18 V
0 17 V
0 18 V
0 18 V
0 18 V
0 18 V
0 17 V
0 18 V
0 18 V
0 18 V
0 17 V
0 18 V
0 18 V
0 18 V
0 18 V
0 17 V
0 18 V
0 18 V
0 18 V
0 17 V
0 18 V
0 18 V
1.000 UL
LTa
3204 300 M
0 18 V
0 18 V
0 17 V
0 18 V
0 18 V
0 18 V
0 17 V
0 18 V
0 18 V
0 18 V
0 18 V
0 17 V
0 18 V
0 18 V
0 18 V
0 17 V
0 18 V
0 18 V
0 18 V
0 18 V
0 17 V
0 18 V
0 18 V
0 18 V
0 17 V
0 18 V
0 18 V
0 18 V
0 18 V
0 17 V
0 18 V
0 18 V
0 18 V
0 17 V
0 18 V
0 18 V
0 18 V
0 18 V
0 17 V
0 18 V
0 18 V
0 18 V
0 17 V
0 18 V
0 18 V
0 18 V
0 18 V
0 17 V
0 18 V
0 18 V
0 18 V
0 17 V
0 18 V
0 18 V
0 18 V
0 18 V
0 17 V
0 18 V
0 18 V
0 18 V
0 17 V
0 18 V
0 18 V
0 18 V
0 18 V
0 17 V
0 18 V
0 18 V
0 18 V
0 17 V
0 18 V
0 18 V
0 18 V
0 18 V
0 17 V
0 18 V
0 18 V
0 18 V
0 17 V
0 18 V
0 18 V
0 18 V
0 18 V
0 17 V
0 18 V
0 18 V
0 18 V
0 17 V
0 18 V
0 18 V
0 18 V
0 18 V
0 17 V
0 18 V
0 18 V
0 18 V
0 17 V
0 18 V
0 18 V
stroke
grestore
end
showpage
}}%
\put(2925,388){\makebox(0,0)[l]{$\alpha_d$}}%
\put(3225,388){\makebox(0,0)[l]{$\alpha_c$}}%
\put(1950,50){\makebox(0,0){$\alpha$}}%
\put(100,1180){%
\special{ps: gsave currentpoint currentpoint translate
270 rotate neg exch neg exch translate}%
\makebox(0,0)[b]{\shortstack{$x_1$}}%
\special{ps: currentpoint grestore moveto}%
}%
\put(3450,200){\makebox(0,0){ 1}}%
\put(3150,200){\makebox(0,0){ 0.9}}%
\put(2850,200){\makebox(0,0){ 0.8}}%
\put(2550,200){\makebox(0,0){ 0.7}}%
\put(2250,200){\makebox(0,0){ 0.6}}%
\put(1950,200){\makebox(0,0){ 0.5}}%
\put(1650,200){\makebox(0,0){ 0.4}}%
\put(1350,200){\makebox(0,0){ 0.3}}%
\put(1050,200){\makebox(0,0){ 0.2}}%
\put(750,200){\makebox(0,0){ 0.1}}%
\put(450,200){\makebox(0,0){ 0}}%
\put(400,2060){\makebox(0,0)[r]{ 1}}%
\put(400,1708){\makebox(0,0)[r]{ 0.8}}%
\put(400,1356){\makebox(0,0)[r]{ 0.6}}%
\put(400,1004){\makebox(0,0)[r]{ 0.4}}%
\put(400,652){\makebox(0,0)[r]{ 0.2}}%
\put(400,300){\makebox(0,0)[r]{ 0}}%
\end{picture}%
\endgroup
 

%% file: clusterdistance.pstex_t
\begin{picture}(0,0)%
\includegraphics{clusterdistance.pstex}%
\end{picture}%
\setlength{\unitlength}{4144sp}%
\begingroup\makeatletter\ifx\SetFigFont\undefined%
\gdef\SetFigFont#1#2#3#4#5{%
  \reset@font\fontsize{#1}{#2pt}%
  \fontfamily{#3}\fontseries{#4}\fontshape{#5}%
  \selectfont}%
\fi\endgroup%
\begin{picture}(3607,1784)(937,-2186)
\put(3919,-815){\makebox(0,0)[lb]{\smash{{\SetFigFont{14}{16.8}{\rmdefault}{\mddefault}{\updefault}{\color[rgb]{0,0,0}$\bc$}%
}}}}
\put(2551,-1451){\makebox(0,0)[lb]{\smash{{\SetFigFont{14}{16.8}{\rmdefault}{\mddefault}{\updefault}{\color[rgb]{0,0,0}$\bze$}%
}}}}
\put(3171,-1191){\makebox(0,0)[lb]{\smash{{\SetFigFont{14}{16.8}{\rmdefault}{\mddefault}{\updefault}{\color[rgb]{0,0,0}$X_\bc$}%
}}}}
\put(3311,-1551){\makebox(0,0)[lb]{\smash{{\SetFigFont{14}{16.8}{\rmdefault}{\mddefault}{\updefault}{\color[rgb]{0,0,0}$x_\bc$}%
}}}}
\end{picture}%

%% file: wefk3.tex
% GNUPLOT: LaTeX picture with Postscript
\begingroup%
  \makeatletter%
  \newcommand{\GNUPLOTspecial}{%
    \@sanitize\catcode`\%=14\relax\special}%
  \setlength{\unitlength}{0.1bp}%
{\GNUPLOTspecial{!
%!PS-Adobe-2.0
%%Title: wefk3.tex
%%Creator: gnuplot 3.7 patchlevel 3
%%CreationDate: Thu Jun  1 11:44:49 2006
%%DocumentFonts: 
%%BoundingBox: 0 0 360 216
%%Orientation: Landscape
%%Pages: (atend)
%%EndComments
/gnudict 256 dict def
gnudict begin
/Color true def
/Solid false def
/gnulinewidth 5.000 def
/userlinewidth gnulinewidth def
/vshift -33 def
/dl {10 mul} def
/hpt_ 31.5 def
/vpt_ 31.5 def
/hpt hpt_ def
/vpt vpt_ def
/M {moveto} bind def
/L {lineto} bind def
/R {rmoveto} bind def
/V {rlineto} bind def
/vpt2 vpt 2 mul def
/hpt2 hpt 2 mul def
/Lshow { currentpoint stroke M
  0 vshift R show } def
/Rshow { currentpoint stroke M
  dup stringwidth pop neg vshift R show } def
/Cshow { currentpoint stroke M
  dup stringwidth pop -2 div vshift R show } def
/UP { dup vpt_ mul /vpt exch def hpt_ mul /hpt exch def
  /hpt2 hpt 2 mul def /vpt2 vpt 2 mul def } def
/DL { Color {setrgbcolor Solid {pop []} if 0 setdash }
 {pop pop pop Solid {pop []} if 0 setdash} ifelse } def
/BL { stroke userlinewidth 2 mul setlinewidth } def
/AL { stroke userlinewidth 2 div setlinewidth } def
/UL { dup gnulinewidth mul /userlinewidth exch def
      dup 1 lt {pop 1} if 10 mul /udl exch def } def
/PL { stroke userlinewidth setlinewidth } def
/LTb { BL [] 0 0 0 DL } def
/LTa { AL [1 udl mul 2 udl mul] 0 setdash 0 0 0 setrgbcolor } def
/LT0 { PL [] 1 0 0 DL } def
/LT1 { PL [4 dl 2 dl] 0 1 0 DL } def
/LT2 { PL [2 dl 3 dl] 0 0 1 DL } def
/LT3 { PL [1 dl 1.5 dl] 1 0 1 DL } def
/LT4 { PL [5 dl 2 dl 1 dl 2 dl] 0 1 1 DL } def
/LT5 { PL [4 dl 3 dl 1 dl 3 dl] 1 1 0 DL } def
/LT6 { PL [2 dl 2 dl 2 dl 4 dl] 0 0 0 DL } def
/LT7 { PL [2 dl 2 dl 2 dl 2 dl 2 dl 4 dl] 1 0.3 0 DL } def
/LT8 { PL [2 dl 2 dl 2 dl 2 dl 2 dl 2 dl 2 dl 4 dl] 0.5 0.5 0.5 DL } def
/Pnt { stroke [] 0 setdash
   gsave 1 setlinecap M 0 0 V stroke grestore } def
/Dia { stroke [] 0 setdash 2 copy vpt add M
  hpt neg vpt neg V hpt vpt neg V
  hpt vpt V hpt neg vpt V closepath stroke
  Pnt } def
/Pls { stroke [] 0 setdash vpt sub M 0 vpt2 V
  currentpoint stroke M
  hpt neg vpt neg R hpt2 0 V stroke
  } def
/Box { stroke [] 0 setdash 2 copy exch hpt sub exch vpt add M
  0 vpt2 neg V hpt2 0 V 0 vpt2 V
  hpt2 neg 0 V closepath stroke
  Pnt } def
/Crs { stroke [] 0 setdash exch hpt sub exch vpt add M
  hpt2 vpt2 neg V currentpoint stroke M
  hpt2 neg 0 R hpt2 vpt2 V stroke } def
/TriU { stroke [] 0 setdash 2 copy vpt 1.12 mul add M
  hpt neg vpt -1.62 mul V
  hpt 2 mul 0 V
  hpt neg vpt 1.62 mul V closepath stroke
  Pnt  } def
/Star { 2 copy Pls Crs } def
/BoxF { stroke [] 0 setdash exch hpt sub exch vpt add M
  0 vpt2 neg V  hpt2 0 V  0 vpt2 V
  hpt2 neg 0 V  closepath fill } def
/TriUF { stroke [] 0 setdash vpt 1.12 mul add M
  hpt neg vpt -1.62 mul V
  hpt 2 mul 0 V
  hpt neg vpt 1.62 mul V closepath fill } def
/TriD { stroke [] 0 setdash 2 copy vpt 1.12 mul sub M
  hpt neg vpt 1.62 mul V
  hpt 2 mul 0 V
  hpt neg vpt -1.62 mul V closepath stroke
  Pnt  } def
/TriDF { stroke [] 0 setdash vpt 1.12 mul sub M
  hpt neg vpt 1.62 mul V
  hpt 2 mul 0 V
  hpt neg vpt -1.62 mul V closepath fill} def
/DiaF { stroke [] 0 setdash vpt add M
  hpt neg vpt neg V hpt vpt neg V
  hpt vpt V hpt neg vpt V closepath fill } def
/Pent { stroke [] 0 setdash 2 copy gsave
  translate 0 hpt M 4 {72 rotate 0 hpt L} repeat
  closepath stroke grestore Pnt } def
/PentF { stroke [] 0 setdash gsave
  translate 0 hpt M 4 {72 rotate 0 hpt L} repeat
  closepath fill grestore } def
/Circle { stroke [] 0 setdash 2 copy
  hpt 0 360 arc stroke Pnt } def
/CircleF { stroke [] 0 setdash hpt 0 360 arc fill } def
/C0 { BL [] 0 setdash 2 copy moveto vpt 90 450  arc } bind def
/C1 { BL [] 0 setdash 2 copy        moveto
       2 copy  vpt 0 90 arc closepath fill
               vpt 0 360 arc closepath } bind def
/C2 { BL [] 0 setdash 2 copy moveto
       2 copy  vpt 90 180 arc closepath fill
               vpt 0 360 arc closepath } bind def
/C3 { BL [] 0 setdash 2 copy moveto
       2 copy  vpt 0 180 arc closepath fill
               vpt 0 360 arc closepath } bind def
/C4 { BL [] 0 setdash 2 copy moveto
       2 copy  vpt 180 270 arc closepath fill
               vpt 0 360 arc closepath } bind def
/C5 { BL [] 0 setdash 2 copy moveto
       2 copy  vpt 0 90 arc
       2 copy moveto
       2 copy  vpt 180 270 arc closepath fill
               vpt 0 360 arc } bind def
/C6 { BL [] 0 setdash 2 copy moveto
      2 copy  vpt 90 270 arc closepath fill
              vpt 0 360 arc closepath } bind def
/C7 { BL [] 0 setdash 2 copy moveto
      2 copy  vpt 0 270 arc closepath fill
              vpt 0 360 arc closepath } bind def
/C8 { BL [] 0 setdash 2 copy moveto
      2 copy vpt 270 360 arc closepath fill
              vpt 0 360 arc closepath } bind def
/C9 { BL [] 0 setdash 2 copy moveto
      2 copy  vpt 270 450 arc closepath fill
              vpt 0 360 arc closepath } bind def
/C10 { BL [] 0 setdash 2 copy 2 copy moveto vpt 270 360 arc closepath fill
       2 copy moveto
       2 copy vpt 90 180 arc closepath fill
               vpt 0 360 arc closepath } bind def
/C11 { BL [] 0 setdash 2 copy moveto
       2 copy  vpt 0 180 arc closepath fill
       2 copy moveto
       2 copy  vpt 270 360 arc closepath fill
               vpt 0 360 arc closepath } bind def
/C12 { BL [] 0 setdash 2 copy moveto
       2 copy  vpt 180 360 arc closepath fill
               vpt 0 360 arc closepath } bind def
/C13 { BL [] 0 setdash  2 copy moveto
       2 copy  vpt 0 90 arc closepath fill
       2 copy moveto
       2 copy  vpt 180 360 arc closepath fill
               vpt 0 360 arc closepath } bind def
/C14 { BL [] 0 setdash 2 copy moveto
       2 copy  vpt 90 360 arc closepath fill
               vpt 0 360 arc } bind def
/C15 { BL [] 0 setdash 2 copy vpt 0 360 arc closepath fill
               vpt 0 360 arc closepath } bind def
/Rec   { newpath 4 2 roll moveto 1 index 0 rlineto 0 exch rlineto
       neg 0 rlineto closepath } bind def
/Square { dup Rec } bind def
/Bsquare { vpt sub exch vpt sub exch vpt2 Square } bind def
/S0 { BL [] 0 setdash 2 copy moveto 0 vpt rlineto BL Bsquare } bind def
/S1 { BL [] 0 setdash 2 copy vpt Square fill Bsquare } bind def
/S2 { BL [] 0 setdash 2 copy exch vpt sub exch vpt Square fill Bsquare } bind def
/S3 { BL [] 0 setdash 2 copy exch vpt sub exch vpt2 vpt Rec fill Bsquare } bind def
/S4 { BL [] 0 setdash 2 copy exch vpt sub exch vpt sub vpt Square fill Bsquare } bind def
/S5 { BL [] 0 setdash 2 copy 2 copy vpt Square fill
       exch vpt sub exch vpt sub vpt Square fill Bsquare } bind def
/S6 { BL [] 0 setdash 2 copy exch vpt sub exch vpt sub vpt vpt2 Rec fill Bsquare } bind def
/S7 { BL [] 0 setdash 2 copy exch vpt sub exch vpt sub vpt vpt2 Rec fill
       2 copy vpt Square fill
       Bsquare } bind def
/S8 { BL [] 0 setdash 2 copy vpt sub vpt Square fill Bsquare } bind def
/S9 { BL [] 0 setdash 2 copy vpt sub vpt vpt2 Rec fill Bsquare } bind def
/S10 { BL [] 0 setdash 2 copy vpt sub vpt Square fill 2 copy exch vpt sub exch vpt Square fill
       Bsquare } bind def
/S11 { BL [] 0 setdash 2 copy vpt sub vpt Square fill 2 copy exch vpt sub exch vpt2 vpt Rec fill
       Bsquare } bind def
/S12 { BL [] 0 setdash 2 copy exch vpt sub exch vpt sub vpt2 vpt Rec fill Bsquare } bind def
/S13 { BL [] 0 setdash 2 copy exch vpt sub exch vpt sub vpt2 vpt Rec fill
       2 copy vpt Square fill Bsquare } bind def
/S14 { BL [] 0 setdash 2 copy exch vpt sub exch vpt sub vpt2 vpt Rec fill
       2 copy exch vpt sub exch vpt Square fill Bsquare } bind def
/S15 { BL [] 0 setdash 2 copy Bsquare fill Bsquare } bind def
/D0 { gsave translate 45 rotate 0 0 S0 stroke grestore } bind def
/D1 { gsave translate 45 rotate 0 0 S1 stroke grestore } bind def
/D2 { gsave translate 45 rotate 0 0 S2 stroke grestore } bind def
/D3 { gsave translate 45 rotate 0 0 S3 stroke grestore } bind def
/D4 { gsave translate 45 rotate 0 0 S4 stroke grestore } bind def
/D5 { gsave translate 45 rotate 0 0 S5 stroke grestore } bind def
/D6 { gsave translate 45 rotate 0 0 S6 stroke grestore } bind def
/D7 { gsave translate 45 rotate 0 0 S7 stroke grestore } bind def
/D8 { gsave translate 45 rotate 0 0 S8 stroke grestore } bind def
/D9 { gsave translate 45 rotate 0 0 S9 stroke grestore } bind def
/D10 { gsave translate 45 rotate 0 0 S10 stroke grestore } bind def
/D11 { gsave translate 45 rotate 0 0 S11 stroke grestore } bind def
/D12 { gsave translate 45 rotate 0 0 S12 stroke grestore } bind def
/D13 { gsave translate 45 rotate 0 0 S13 stroke grestore } bind def
/D14 { gsave translate 45 rotate 0 0 S14 stroke grestore } bind def
/D15 { gsave translate 45 rotate 0 0 S15 stroke grestore } bind def
/DiaE { stroke [] 0 setdash vpt add M
  hpt neg vpt neg V hpt vpt neg V
  hpt vpt V hpt neg vpt V closepath stroke } def
/BoxE { stroke [] 0 setdash exch hpt sub exch vpt add M
  0 vpt2 neg V hpt2 0 V 0 vpt2 V
  hpt2 neg 0 V closepath stroke } def
/TriUE { stroke [] 0 setdash vpt 1.12 mul add M
  hpt neg vpt -1.62 mul V
  hpt 2 mul 0 V
  hpt neg vpt 1.62 mul V closepath stroke } def
/TriDE { stroke [] 0 setdash vpt 1.12 mul sub M
  hpt neg vpt 1.62 mul V
  hpt 2 mul 0 V
  hpt neg vpt -1.62 mul V closepath stroke } def
/PentE { stroke [] 0 setdash gsave
  translate 0 hpt M 4 {72 rotate 0 hpt L} repeat
  closepath stroke grestore } def
/CircE { stroke [] 0 setdash 
  hpt 0 360 arc stroke } def
/Opaque { gsave closepath 1 setgray fill grestore 0 setgray closepath } def
/DiaW { stroke [] 0 setdash vpt add M
  hpt neg vpt neg V hpt vpt neg V
  hpt vpt V hpt neg vpt V Opaque stroke } def
/BoxW { stroke [] 0 setdash exch hpt sub exch vpt add M
  0 vpt2 neg V hpt2 0 V 0 vpt2 V
  hpt2 neg 0 V Opaque stroke } def
/TriUW { stroke [] 0 setdash vpt 1.12 mul add M
  hpt neg vpt -1.62 mul V
  hpt 2 mul 0 V
  hpt neg vpt 1.62 mul V Opaque stroke } def
/TriDW { stroke [] 0 setdash vpt 1.12 mul sub M
  hpt neg vpt 1.62 mul V
  hpt 2 mul 0 V
  hpt neg vpt -1.62 mul V Opaque stroke } def
/PentW { stroke [] 0 setdash gsave
  translate 0 hpt M 4 {72 rotate 0 hpt L} repeat
  Opaque stroke grestore } def
/CircW { stroke [] 0 setdash 
  hpt 0 360 arc Opaque stroke } def
/BoxFill { gsave Rec 1 setgray fill grestore } def
/Symbol-Oblique /Symbol findfont [1 0 .167 1 0 0] makefont
dup length dict begin {1 index /FID eq {pop pop} {def} ifelse} forall
currentdict end definefont pop
end
}}%
\begin{picture}(3600,2160)(0,0)%
{\GNUPLOTspecial{"
gnudict begin
gsave
0 0 translate
0.100 0.100 scale
0 setgray
newpath
1.000 UL
LTb
550 300 M
63 0 V
2837 0 R
-63 0 V
550 520 M
63 0 V
2837 0 R
-63 0 V
550 740 M
63 0 V
2837 0 R
-63 0 V
550 960 M
63 0 V
2837 0 R
-63 0 V
550 1180 M
63 0 V
2837 0 R
-63 0 V
550 1400 M
63 0 V
2837 0 R
-63 0 V
550 1620 M
63 0 V
2837 0 R
-63 0 V
550 1840 M
63 0 V
2837 0 R
-63 0 V
550 2060 M
63 0 V
2837 0 R
-63 0 V
550 300 M
0 63 V
0 1697 R
0 -63 V
913 300 M
0 63 V
0 1697 R
0 -63 V
1275 300 M
0 63 V
0 1697 R
0 -63 V
1638 300 M
0 63 V
0 1697 R
0 -63 V
2000 300 M
0 63 V
0 1697 R
0 -63 V
2363 300 M
0 63 V
0 1697 R
0 -63 V
2725 300 M
0 63 V
0 1697 R
0 -63 V
3087 300 M
0 63 V
0 1697 R
0 -63 V
3450 300 M
0 63 V
0 1697 R
0 -63 V
1.000 UL
LTa
550 740 M
2900 0 V
1.000 UL
LTb
550 300 M
2900 0 V
0 1760 V
-2900 0 V
550 300 L
0.500 UP
1.000 UL
LT0
1130 410 Pls
2155 1855 Pls
2155 1855 Pls
1166 411 Pls
1094 413 Pls
2146 1855 Pls
2164 1855 Pls
1202 417 Pls
2137 1853 Pls
2171 1854 Pls
2128 1851 Pls
1239 428 Pls
2118 1846 Pls
2187 1848 Pls
1275 445 Pls
2108 1841 Pls
2194 1844 Pls
1311 467 Pls
2099 1836 Pls
2202 1838 Pls
1347 494 Pls
2088 1828 Pls
2209 1834 Pls
2077 1818 Pls
2216 1827 Pls
1384 527 Pls
2065 1805 Pls
2223 1820 Pls
1420 565 Pls
2053 1790 Pls
2229 1813 Pls
1456 610 Pls
2236 1805 Pls
2242 1796 Pls
1492 660 Pls
2249 1786 Pls
1529 717 Pls
2000 1713 Pls
2255 1776 Pls
1565 779 Pls
2261 1765 Pls
1964 1650 Pls
1601 847 Pls
2267 1754 Pls
1637 919 Pls
2272 1743 Pls
1927 1575 Pls
1674 997 Pls
2277 1730 Pls
1891 1496 Pls
1710 1076 Pls
1855 1416 Pls
2283 1718 Pls
1746 1156 Pls
1819 1329 Pls
1782 1242 Pls
2290 1704 Pls
2295 1690 Pls
2301 1676 Pls
2306 1661 Pls
2310 1648 Pls
2317 1630 Pls
2322 1613 Pls
2328 1597 Pls
2331 1583 Pls
2335 1567 Pls
2341 1547 Pls
2348 1526 Pls
2351 1510 Pls
2356 1494 Pls
2359 1478 Pls
2367 1452 Pls
2367 1441 Pls
2375 1414 Pls
2377 1400 Pls
2385 1373 Pls
2386 1357 Pls
2391 1335 Pls
2394 1316 Pls
2400 1291 Pls
2404 1273 Pls
2406 1251 Pls
2408 1243 Pls
2412 1214 Pls
2417 1192 Pls
2421 1168 Pls
2426 1146 Pls
2425 1135 Pls
2430 1109 Pls
2435 1083 Pls
2437 1063 Pls
2442 1039 Pls
2443 1024 Pls
2451 985 Pls
2451 976 Pls
2456 942 Pls
2455 935 Pls
2463 892 Pls
2463 882 Pls
2470 847 Pls
2471 831 Pls
2474 807 Pls
2476 789 Pls
2478 767 Pls
2483 734 Pls
2487 708 Pls
2494 661 Pls
2491 667 Pls
2490 665 Pls
2493 639 Pls
2501 596 Pls
2500 583 Pls
2502 566 Pls
2501 569 Pls
2507 522 Pls
2509 506 Pls
2518 441 Pls
2515 450 Pls
2517 423 Pls
2523 392 Pls
2522 388 Pls
2527 346 Pls
2528 328 Pls
2531 312 Pls
-2147483648 -2147483648 Pls
-2147483648 -2147483648 Pls
3234 1947 Pls
0.500 UP
1.000 UL
LT1
2574 1855 Crs
2559 1855 Crs
2580 1854 Crs
2550 1854 Crs
2588 1852 Crs
2544 1851 Crs
2594 1848 Crs
2534 1848 Crs
2601 1845 Crs
2526 1844 Crs
2608 1840 Crs
2519 1838 Crs
2613 1836 Crs
2508 1832 Crs
2620 1830 Crs
2499 1823 Crs
2626 1824 Crs
2490 1813 Crs
2631 1817 Crs
2479 1800 Crs
2637 1811 Crs
2468 1786 Crs
2643 1803 Crs
2458 1771 Crs
2649 1795 Crs
2445 1753 Crs
2654 1784 Crs
2432 1731 Crs
2660 1775 Crs
2419 1707 Crs
2666 1765 Crs
2404 1679 Crs
2671 1755 Crs
2387 1646 Crs
2676 1743 Crs
2369 1605 Crs
2681 1732 Crs
2346 1554 Crs
2686 1720 Crs
2319 1489 Crs
2691 1707 Crs
2275 1381 Crs
2696 1695 Crs
2701 1681 Crs
2706 1667 Crs
2710 1653 Crs
2715 1639 Crs
2720 1624 Crs
2724 1609 Crs
2729 1593 Crs
2733 1578 Crs
2737 1560 Crs
2742 1544 Crs
2747 1527 Crs
2751 1509 Crs
2755 1491 Crs
2759 1473 Crs
2763 1455 Crs
2768 1436 Crs
2771 1418 Crs
2775 1398 Crs
2779 1379 Crs
2783 1358 Crs
2787 1338 Crs
2791 1318 Crs
2795 1297 Crs
2799 1278 Crs
2803 1255 Crs
2806 1234 Crs
2810 1211 Crs
2814 1189 Crs
2817 1167 Crs
2821 1145 Crs
2825 1122 Crs
2828 1098 Crs
2832 1075 Crs
2835 1051 Crs
2839 1026 Crs
2842 1005 Crs
2846 980 Crs
2849 954 Crs
2852 930 Crs
2856 905 Crs
2859 881 Crs
2862 854 Crs
2866 830 Crs
2869 803 Crs
2872 774 Crs
2875 749 Crs
2878 724 Crs
2882 695 Crs
2885 669 Crs
2888 645 Crs
2891 617 Crs
2894 591 Crs
2897 562 Crs
2900 534 Crs
2903 506 Crs
2906 477 Crs
2909 449 Crs
2912 421 Crs
2915 393 Crs
2918 366 Crs
2921 337 Crs
2923 307 Crs
3234 1847 Crs
stroke
grestore
end
showpage
}}%
\put(3069,1847){\makebox(0,0)[r]{$\Sigma_M(x)$}}%
\put(3069,1947){\makebox(0,0)[r]{$\Sigma_m(x)$}}%
\put(2000,50){\makebox(0,0){$x$}}%
\put(100,1180){%
\special{ps: gsave currentpoint currentpoint translate
270 rotate neg exch neg exch translate}%
\makebox(0,0)[b]{\shortstack{$\Sigma_m(x),\Sigma_M(x)$}}%
\special{ps: currentpoint grestore moveto}%
}%
\put(3450,200){\makebox(0,0){ 0.8}}%
\put(3087,200){\makebox(0,0){ 0.7}}%
\put(2725,200){\makebox(0,0){ 0.6}}%
\put(2363,200){\makebox(0,0){ 0.5}}%
\put(2000,200){\makebox(0,0){ 0.4}}%
\put(1638,200){\makebox(0,0){ 0.3}}%
\put(1275,200){\makebox(0,0){ 0.2}}%
\put(913,200){\makebox(0,0){ 0.1}}%
\put(550,200){\makebox(0,0){ 0}}%
\put(500,2060){\makebox(0,0)[r]{ 0.03}}%
\put(500,1840){\makebox(0,0)[r]{ 0.025}}%
\put(500,1620){\makebox(0,0)[r]{ 0.02}}%
\put(500,1400){\makebox(0,0)[r]{ 0.015}}%
\put(500,1180){\makebox(0,0)[r]{ 0.01}}%
\put(500,960){\makebox(0,0)[r]{ 0.005}}%
\put(500,740){\makebox(0,0)[r]{ 0}}%
\put(500,520){\makebox(0,0)[r]{-0.005}}%
\put(500,300){\makebox(0,0)[r]{-0.01}}%
\end{picture}%
\endgroup
 

%% file: alphaxphasek3.tex
% GNUPLOT: LaTeX picture with Postscript
\begingroup%
  \makeatletter%
  \newcommand{\GNUPLOTspecial}{%
    \@sanitize\catcode`\%=14\relax\special}%
  \setlength{\unitlength}{0.1bp}%
{\GNUPLOTspecial{!
%!PS-Adobe-2.0
%%Title: alphaxphasek3.tex
%%Creator: gnuplot 3.7 patchlevel 3
%%CreationDate: Mon May 15 20:13:16 2006
%%DocumentFonts: 
%%BoundingBox: 0 0 360 216
%%Orientation: Landscape
%%Pages: (atend)
%%EndComments
/gnudict 256 dict def
gnudict begin
/Color true def
/Solid false def
/gnulinewidth 5.000 def
/userlinewidth gnulinewidth def
/vshift -33 def
/dl {10 mul} def
/hpt_ 31.5 def
/vpt_ 31.5 def
/hpt hpt_ def
/vpt vpt_ def
/M {moveto} bind def
/L {lineto} bind def
/R {rmoveto} bind def
/V {rlineto} bind def
/vpt2 vpt 2 mul def
/hpt2 hpt 2 mul def
/Lshow { currentpoint stroke M
  0 vshift R show } def
/Rshow { currentpoint stroke M
  dup stringwidth pop neg vshift R show } def
/Cshow { currentpoint stroke M
  dup stringwidth pop -2 div vshift R show } def
/UP { dup vpt_ mul /vpt exch def hpt_ mul /hpt exch def
  /hpt2 hpt 2 mul def /vpt2 vpt 2 mul def } def
/DL { Color {setrgbcolor Solid {pop []} if 0 setdash }
 {pop pop pop Solid {pop []} if 0 setdash} ifelse } def
/BL { stroke userlinewidth 2 mul setlinewidth } def
/AL { stroke userlinewidth 2 div setlinewidth } def
/UL { dup gnulinewidth mul /userlinewidth exch def
      dup 1 lt {pop 1} if 10 mul /udl exch def } def
/PL { stroke userlinewidth setlinewidth } def
/LTb { BL [] 0 0 0 DL } def
/LTa { AL [1 udl mul 2 udl mul] 0 setdash 0 0 0 setrgbcolor } def
/LT0 { PL [] 1 0 0 DL } def
/LT1 { PL [4 dl 2 dl] 0 1 0 DL } def
/LT2 { PL [2 dl 3 dl] 0 0 1 DL } def
/LT3 { PL [1 dl 1.5 dl] 1 0 1 DL } def
/LT4 { PL [5 dl 2 dl 1 dl 2 dl] 0 1 1 DL } def
/LT5 { PL [4 dl 3 dl 1 dl 3 dl] 1 1 0 DL } def
/LT6 { PL [2 dl 2 dl 2 dl 4 dl] 0 0 0 DL } def
/LT7 { PL [2 dl 2 dl 2 dl 2 dl 2 dl 4 dl] 1 0.3 0 DL } def
/LT8 { PL [2 dl 2 dl 2 dl 2 dl 2 dl 2 dl 2 dl 4 dl] 0.5 0.5 0.5 DL } def
/Pnt { stroke [] 0 setdash
   gsave 1 setlinecap M 0 0 V stroke grestore } def
/Dia { stroke [] 0 setdash 2 copy vpt add M
  hpt neg vpt neg V hpt vpt neg V
  hpt vpt V hpt neg vpt V closepath stroke
  Pnt } def
/Pls { stroke [] 0 setdash vpt sub M 0 vpt2 V
  currentpoint stroke M
  hpt neg vpt neg R hpt2 0 V stroke
  } def
/Box { stroke [] 0 setdash 2 copy exch hpt sub exch vpt add M
  0 vpt2 neg V hpt2 0 V 0 vpt2 V
  hpt2 neg 0 V closepath stroke
  Pnt } def
/Crs { stroke [] 0 setdash exch hpt sub exch vpt add M
  hpt2 vpt2 neg V currentpoint stroke M
  hpt2 neg 0 R hpt2 vpt2 V stroke } def
/TriU { stroke [] 0 setdash 2 copy vpt 1.12 mul add M
  hpt neg vpt -1.62 mul V
  hpt 2 mul 0 V
  hpt neg vpt 1.62 mul V closepath stroke
  Pnt  } def
/Star { 2 copy Pls Crs } def
/BoxF { stroke [] 0 setdash exch hpt sub exch vpt add M
  0 vpt2 neg V  hpt2 0 V  0 vpt2 V
  hpt2 neg 0 V  closepath fill } def
/TriUF { stroke [] 0 setdash vpt 1.12 mul add M
  hpt neg vpt -1.62 mul V
  hpt 2 mul 0 V
  hpt neg vpt 1.62 mul V closepath fill } def
/TriD { stroke [] 0 setdash 2 copy vpt 1.12 mul sub M
  hpt neg vpt 1.62 mul V
  hpt 2 mul 0 V
  hpt neg vpt -1.62 mul V closepath stroke
  Pnt  } def
/TriDF { stroke [] 0 setdash vpt 1.12 mul sub M
  hpt neg vpt 1.62 mul V
  hpt 2 mul 0 V
  hpt neg vpt -1.62 mul V closepath fill} def
/DiaF { stroke [] 0 setdash vpt add M
  hpt neg vpt neg V hpt vpt neg V
  hpt vpt V hpt neg vpt V closepath fill } def
/Pent { stroke [] 0 setdash 2 copy gsave
  translate 0 hpt M 4 {72 rotate 0 hpt L} repeat
  closepath stroke grestore Pnt } def
/PentF { stroke [] 0 setdash gsave
  translate 0 hpt M 4 {72 rotate 0 hpt L} repeat
  closepath fill grestore } def
/Circle { stroke [] 0 setdash 2 copy
  hpt 0 360 arc stroke Pnt } def
/CircleF { stroke [] 0 setdash hpt 0 360 arc fill } def
/C0 { BL [] 0 setdash 2 copy moveto vpt 90 450  arc } bind def
/C1 { BL [] 0 setdash 2 copy        moveto
       2 copy  vpt 0 90 arc closepath fill
               vpt 0 360 arc closepath } bind def
/C2 { BL [] 0 setdash 2 copy moveto
       2 copy  vpt 90 180 arc closepath fill
               vpt 0 360 arc closepath } bind def
/C3 { BL [] 0 setdash 2 copy moveto
       2 copy  vpt 0 180 arc closepath fill
               vpt 0 360 arc closepath } bind def
/C4 { BL [] 0 setdash 2 copy moveto
       2 copy  vpt 180 270 arc closepath fill
               vpt 0 360 arc closepath } bind def
/C5 { BL [] 0 setdash 2 copy moveto
       2 copy  vpt 0 90 arc
       2 copy moveto
       2 copy  vpt 180 270 arc closepath fill
               vpt 0 360 arc } bind def
/C6 { BL [] 0 setdash 2 copy moveto
      2 copy  vpt 90 270 arc closepath fill
              vpt 0 360 arc closepath } bind def
/C7 { BL [] 0 setdash 2 copy moveto
      2 copy  vpt 0 270 arc closepath fill
              vpt 0 360 arc closepath } bind def
/C8 { BL [] 0 setdash 2 copy moveto
      2 copy vpt 270 360 arc closepath fill
              vpt 0 360 arc closepath } bind def
/C9 { BL [] 0 setdash 2 copy moveto
      2 copy  vpt 270 450 arc closepath fill
              vpt 0 360 arc closepath } bind def
/C10 { BL [] 0 setdash 2 copy 2 copy moveto vpt 270 360 arc closepath fill
       2 copy moveto
       2 copy vpt 90 180 arc closepath fill
               vpt 0 360 arc closepath } bind def
/C11 { BL [] 0 setdash 2 copy moveto
       2 copy  vpt 0 180 arc closepath fill
       2 copy moveto
       2 copy  vpt 270 360 arc closepath fill
               vpt 0 360 arc closepath } bind def
/C12 { BL [] 0 setdash 2 copy moveto
       2 copy  vpt 180 360 arc closepath fill
               vpt 0 360 arc closepath } bind def
/C13 { BL [] 0 setdash  2 copy moveto
       2 copy  vpt 0 90 arc closepath fill
       2 copy moveto
       2 copy  vpt 180 360 arc closepath fill
               vpt 0 360 arc closepath } bind def
/C14 { BL [] 0 setdash 2 copy moveto
       2 copy  vpt 90 360 arc closepath fill
               vpt 0 360 arc } bind def
/C15 { BL [] 0 setdash 2 copy vpt 0 360 arc closepath fill
               vpt 0 360 arc closepath } bind def
/Rec   { newpath 4 2 roll moveto 1 index 0 rlineto 0 exch rlineto
       neg 0 rlineto closepath } bind def
/Square { dup Rec } bind def
/Bsquare { vpt sub exch vpt sub exch vpt2 Square } bind def
/S0 { BL [] 0 setdash 2 copy moveto 0 vpt rlineto BL Bsquare } bind def
/S1 { BL [] 0 setdash 2 copy vpt Square fill Bsquare } bind def
/S2 { BL [] 0 setdash 2 copy exch vpt sub exch vpt Square fill Bsquare } bind def
/S3 { BL [] 0 setdash 2 copy exch vpt sub exch vpt2 vpt Rec fill Bsquare } bind def
/S4 { BL [] 0 setdash 2 copy exch vpt sub exch vpt sub vpt Square fill Bsquare } bind def
/S5 { BL [] 0 setdash 2 copy 2 copy vpt Square fill
       exch vpt sub exch vpt sub vpt Square fill Bsquare } bind def
/S6 { BL [] 0 setdash 2 copy exch vpt sub exch vpt sub vpt vpt2 Rec fill Bsquare } bind def
/S7 { BL [] 0 setdash 2 copy exch vpt sub exch vpt sub vpt vpt2 Rec fill
       2 copy vpt Square fill
       Bsquare } bind def
/S8 { BL [] 0 setdash 2 copy vpt sub vpt Square fill Bsquare } bind def
/S9 { BL [] 0 setdash 2 copy vpt sub vpt vpt2 Rec fill Bsquare } bind def
/S10 { BL [] 0 setdash 2 copy vpt sub vpt Square fill 2 copy exch vpt sub exch vpt Square fill
       Bsquare } bind def
/S11 { BL [] 0 setdash 2 copy vpt sub vpt Square fill 2 copy exch vpt sub exch vpt2 vpt Rec fill
       Bsquare } bind def
/S12 { BL [] 0 setdash 2 copy exch vpt sub exch vpt sub vpt2 vpt Rec fill Bsquare } bind def
/S13 { BL [] 0 setdash 2 copy exch vpt sub exch vpt sub vpt2 vpt Rec fill
       2 copy vpt Square fill Bsquare } bind def
/S14 { BL [] 0 setdash 2 copy exch vpt sub exch vpt sub vpt2 vpt Rec fill
       2 copy exch vpt sub exch vpt Square fill Bsquare } bind def
/S15 { BL [] 0 setdash 2 copy Bsquare fill Bsquare } bind def
/D0 { gsave translate 45 rotate 0 0 S0 stroke grestore } bind def
/D1 { gsave translate 45 rotate 0 0 S1 stroke grestore } bind def
/D2 { gsave translate 45 rotate 0 0 S2 stroke grestore } bind def
/D3 { gsave translate 45 rotate 0 0 S3 stroke grestore } bind def
/D4 { gsave translate 45 rotate 0 0 S4 stroke grestore } bind def
/D5 { gsave translate 45 rotate 0 0 S5 stroke grestore } bind def
/D6 { gsave translate 45 rotate 0 0 S6 stroke grestore } bind def
/D7 { gsave translate 45 rotate 0 0 S7 stroke grestore } bind def
/D8 { gsave translate 45 rotate 0 0 S8 stroke grestore } bind def
/D9 { gsave translate 45 rotate 0 0 S9 stroke grestore } bind def
/D10 { gsave translate 45 rotate 0 0 S10 stroke grestore } bind def
/D11 { gsave translate 45 rotate 0 0 S11 stroke grestore } bind def
/D12 { gsave translate 45 rotate 0 0 S12 stroke grestore } bind def
/D13 { gsave translate 45 rotate 0 0 S13 stroke grestore } bind def
/D14 { gsave translate 45 rotate 0 0 S14 stroke grestore } bind def
/D15 { gsave translate 45 rotate 0 0 S15 stroke grestore } bind def
/DiaE { stroke [] 0 setdash vpt add M
  hpt neg vpt neg V hpt vpt neg V
  hpt vpt V hpt neg vpt V closepath stroke } def
/BoxE { stroke [] 0 setdash exch hpt sub exch vpt add M
  0 vpt2 neg V hpt2 0 V 0 vpt2 V
  hpt2 neg 0 V closepath stroke } def
/TriUE { stroke [] 0 setdash vpt 1.12 mul add M
  hpt neg vpt -1.62 mul V
  hpt 2 mul 0 V
  hpt neg vpt 1.62 mul V closepath stroke } def
/TriDE { stroke [] 0 setdash vpt 1.12 mul sub M
  hpt neg vpt 1.62 mul V
  hpt 2 mul 0 V
  hpt neg vpt -1.62 mul V closepath stroke } def
/PentE { stroke [] 0 setdash gsave
  translate 0 hpt M 4 {72 rotate 0 hpt L} repeat
  closepath stroke grestore } def
/CircE { stroke [] 0 setdash 
  hpt 0 360 arc stroke } def
/Opaque { gsave closepath 1 setgray fill grestore 0 setgray closepath } def
/DiaW { stroke [] 0 setdash vpt add M
  hpt neg vpt neg V hpt vpt neg V
  hpt vpt V hpt neg vpt V Opaque stroke } def
/BoxW { stroke [] 0 setdash exch hpt sub exch vpt add M
  0 vpt2 neg V hpt2 0 V 0 vpt2 V
  hpt2 neg 0 V Opaque stroke } def
/TriUW { stroke [] 0 setdash vpt 1.12 mul add M
  hpt neg vpt -1.62 mul V
  hpt 2 mul 0 V
  hpt neg vpt 1.62 mul V Opaque stroke } def
/TriDW { stroke [] 0 setdash vpt 1.12 mul sub M
  hpt neg vpt 1.62 mul V
  hpt 2 mul 0 V
  hpt neg vpt -1.62 mul V Opaque stroke } def
/PentW { stroke [] 0 setdash gsave
  translate 0 hpt M 4 {72 rotate 0 hpt L} repeat
  Opaque stroke grestore } def
/CircW { stroke [] 0 setdash 
  hpt 0 360 arc Opaque stroke } def
/BoxFill { gsave Rec 1 setgray fill grestore } def
/Symbol-Oblique /Symbol findfont [1 0 .167 1 0 0] makefont
dup length dict begin {1 index /FID eq {pop pop} {def} ifelse} forall
currentdict end definefont pop
end
}}%
\begin{picture}(3600,2160)(0,0)%
{\GNUPLOTspecial{"
gnudict begin
gsave
0 0 translate
0.100 0.100 scale
0 setgray
newpath
1.000 UL
LTb
500 300 M
63 0 V
2887 0 R
-63 0 V
500 652 M
63 0 V
2887 0 R
-63 0 V
500 1004 M
63 0 V
2887 0 R
-63 0 V
500 1356 M
63 0 V
2887 0 R
-63 0 V
500 1708 M
63 0 V
2887 0 R
-63 0 V
500 2060 M
63 0 V
2887 0 R
-63 0 V
500 300 M
0 63 V
0 1697 R
0 -63 V
869 300 M
0 63 V
0 1697 R
0 -63 V
1238 300 M
0 63 V
0 1697 R
0 -63 V
1606 300 M
0 63 V
0 1697 R
0 -63 V
1975 300 M
0 63 V
0 1697 R
0 -63 V
2344 300 M
0 63 V
0 1697 R
0 -63 V
2712 300 M
0 63 V
0 1697 R
0 -63 V
3081 300 M
0 63 V
0 1697 R
0 -63 V
3450 300 M
0 63 V
0 1697 R
0 -63 V
1.000 UL
LTb
500 300 M
2950 0 V
0 1760 V
-2950 0 V
500 300 L
0.500 UP
1.000 UL
LT0
500 1134 Pls
638 1145 Pls
964 1180 Pls
1107 1215 Pls
1232 1286 Pls
1393 1356 Pls
1506 1426 Pls
1613 1497 Pls
1690 1567 Pls
1776 1638 Pls
1908 1708 Pls
2014 1778 Pls
2055 1814 Pls
2182 1834 Pls
1.000 UL
LT0
500 1137 M
17 0 V
17 0 V
17 1 V
17 0 V
17 1 V
17 0 V
17 1 V
17 1 V
17 1 V
17 1 V
17 1 V
17 1 V
17 2 V
17 1 V
17 1 V
17 2 V
17 2 V
17 2 V
17 2 V
17 2 V
17 3 V
17 2 V
17 3 V
17 3 V
17 3 V
17 4 V
17 3 V
17 4 V
17 4 V
17 4 V
17 5 V
17 5 V
17 5 V
17 5 V
17 6 V
17 6 V
17 6 V
17 6 V
17 7 V
17 7 V
17 7 V
17 8 V
17 7 V
17 8 V
17 8 V
17 9 V
17 9 V
17 9 V
17 9 V
17 9 V
17 10 V
17 10 V
17 10 V
17 11 V
17 10 V
17 11 V
17 11 V
17 11 V
17 11 V
17 12 V
17 11 V
17 12 V
17 12 V
17 12 V
16 12 V
17 12 V
17 13 V
17 12 V
17 13 V
17 12 V
17 13 V
17 12 V
17 13 V
17 12 V
17 13 V
17 12 V
17 13 V
17 12 V
17 13 V
17 12 V
17 12 V
17 11 V
17 12 V
17 11 V
17 11 V
17 11 V
17 10 V
17 10 V
17 9 V
17 9 V
17 8 V
17 8 V
17 7 V
17 6 V
17 5 V
17 4 V
17 3 V
17 2 V
17 1 V
0.500 UP
1.000 UL
LT2
2986 1135 Crs
2977 1139 Crs
3001 1146 Crs
3002 1145 Crs
2978 1155 Crs
2977 1159 Crs
2952 1201 Crs
2939 1243 Crs
2927 1286 Crs
2908 1328 Crs
2891 1370 Crs
2875 1412 Crs
2858 1455 Crs
2835 1497 Crs
2817 1539 Crs
2799 1581 Crs
2770 1624 Crs
2742 1666 Crs
2699 1722 Crs
2677 1750 Crs
2615 1792 Crs
2460 1834 Crs
1.000 UL
LT3
500 1828 M
426 0 V
2 -7 V
2 -7 V
3 -7 V
2 -7 V
2 -8 V
3 -7 V
2 -7 V
2 -7 V
3 -7 V
2 -7 V
3 -7 V
2 -7 V
3 -7 V
2 -7 V
3 -7 V
2 -7 V
3 -7 V
3 -7 V
2 -7 V
3 -7 V
2 -7 V
3 -7 V
3 -7 V
3 -7 V
3 -7 V
2 -7 V
3 -7 V
3 -7 V
3 -7 V
3 -8 V
3 -7 V
3 -7 V
3 -7 V
3 -7 V
4 -7 V
2 -7 V
4 -7 V
3 -7 V
3 -7 V
3 -7 V
4 -7 V
3 -7 V
4 -7 V
3 -7 V
4 -7 V
3 -7 V
4 -7 V
3 -7 V
4 -7 V
4 -7 V
4 -7 V
3 -7 V
4 -7 V
4 -7 V
4 -8 V
4 -7 V
4 -7 V
5 -7 V
4 -7 V
4 -7 V
5 -7 V
4 -7 V
4 -7 V
5 -7 V
5 -7 V
4 -7 V
5 -7 V
5 -7 V
6 -7 V
5 -7 V
5 -7 V
5 -7 V
5 -7 V
6 -7 V
6 -7 V
5 -7 V
6 -7 V
7 -7 V
6 -7 V
6 -8 V
7 -7 V
7 -7 V
7 -7 V
7 -7 V
7 -7 V
9 -7 V
7 -7 V
10 -7 V
8 -7 V
9 -7 V
11 -7 V
10 -7 V
12 -7 V
12 -7 V
13 -7 V
16 -7 V
20 -7 V
24 -7 V
1608 -7 V
1 -7 V
0 -7 V
1 -7 V
1 -7 V
1 -7 V
1 -8 V
1 -7 V
1 -7 V
1 -7 V
0 -7 V
1 -7 V
1 -7 V
1 -7 V
1 -7 V
0 -7 V
1 -7 V
1 -7 V
1 -7 V
1 -7 V
1 -7 V
1 -7 V
1 -7 V
0 -7 V
2 -7 V
0 -7 V
1 -7 V
1 -7 V
1 -7 V
1 -7 V
1 -7 V
0 -8 V
1 -7 V
1 -7 V
1 -7 V
1 -7 V
1 -7 V
0 -7 V
1 -7 V
1 -7 V
1 -7 V
1 -7 V
1 -7 V
0 -7 V
1 -7 V
1 -7 V
1 -7 V
1 -7 V
1 -7 V
0 -7 V
1 -7 V
1 -7 V
1 -7 V
0 -7 V
2 -7 V
0 -7 V
1 -8 V
1 -7 V
1 -7 V
1 -7 V
0 -7 V
1 -7 V
1 -7 V
1 -7 V
1 -7 V
0 -7 V
1 -7 V
1 -7 V
1 -7 V
1 -7 V
0 -7 V
1 -7 V
1 -7 V
1 -7 V
1 -7 V
1 -7 V
1 -7 V
0 -7 V
1 -7 V
1 -7 V
1 -7 V
1 -8 V
1 -7 V
0 -7 V
1 -7 V
1 -7 V
1 -7 V
0 -7 V
1 -7 V
1 -7 V
1 -7 V
1 -7 V
1 -7 V
0 -7 V
1 -7 V
1 -7 V
1 -7 V
0 -7 V
1 -7 V
1 -7 V
1 -7 V
1 -7 V
0 -7 V
1 -7 V
1 -7 V
1 -7 V
1 -8 V
1 -7 V
0 -7 V
1 -7 V
1 -7 V
1 -7 V
1 -7 V
0 -7 V
1 -7 V
1 -7 V
1 -7 V
1 -7 V
0.500 UP
1.000 UL
LT3
926 1828 Box
937 1792 Box
949 1757 Box
962 1722 Box
975 1687 Box
989 1652 Box
1003 1616 Box
1019 1581 Box
1034 1546 Box
1052 1511 Box
1070 1476 Box
1089 1440 Box
1110 1405 Box
1133 1370 Box
1158 1335 Box
1185 1300 Box
1215 1264 Box
1250 1229 Box
1293 1194 Box
1351 1159 Box
3020 1124 Box
3024 1088 Box
3028 1053 Box
3032 1018 Box
3037 983 Box
3041 948 Box
3045 912 Box
3050 877 Box
3054 842 Box
3058 807 Box
3062 772 Box
3066 736 Box
3070 701 Box
3074 666 Box
3078 631 Box
3083 596 Box
3087 560 Box
3091 525 Box
3095 490 Box
3099 455 Box
3103 420 Box
3107 384 Box
3111 349 Box
3115 314 Box
1.000 UL
LTb
500 1828 M
426 0 V
2 -7 V
2 -7 V
3 -7 V
2 -7 V
2 -8 V
3 -7 V
2 -7 V
2 -7 V
3 -7 V
2 -7 V
3 -7 V
2 -7 V
3 -7 V
2 -7 V
3 -7 V
2 -7 V
3 -7 V
3 -7 V
2 -7 V
3 -7 V
2 -7 V
3 -7 V
3 -7 V
3 -7 V
3 -7 V
2 -7 V
3 -7 V
3 -7 V
3 -7 V
3 -8 V
3 -7 V
3 -7 V
3 -7 V
3 -7 V
4 -7 V
2 -7 V
4 -7 V
3 -7 V
3 -7 V
3 -7 V
4 -7 V
3 -7 V
4 -7 V
3 -7 V
4 -7 V
3 -7 V
4 -7 V
3 -7 V
4 -7 V
4 -7 V
4 -7 V
3 -7 V
4 -7 V
4 -7 V
4 -8 V
4 -7 V
4 -7 V
5 -7 V
4 -7 V
4 -7 V
5 -7 V
4 -7 V
4 -7 V
5 -7 V
5 -7 V
4 -7 V
5 -7 V
5 -7 V
6 -7 V
5 -7 V
5 -7 V
5 -7 V
5 -7 V
6 -7 V
6 -7 V
5 -7 V
6 -7 V
7 -7 V
6 -7 V
6 -8 V
1.000 UL
LTb
1219 1267 M
10 4 V
10 5 V
9 5 V
10 4 V
10 5 V
9 5 V
10 5 V
10 5 V
10 5 V
9 5 V
10 5 V
10 6 V
10 5 V
9 6 V
10 5 V
10 6 V
9 6 V
10 5 V
10 6 V
10 6 V
9 6 V
10 6 V
10 6 V
10 7 V
9 6 V
10 6 V
10 7 V
9 6 V
10 6 V
10 7 V
10 7 V
9 6 V
10 7 V
10 7 V
10 6 V
9 7 V
10 7 V
10 7 V
9 7 V
10 7 V
10 7 V
10 7 V
9 7 V
10 7 V
10 7 V
10 7 V
9 7 V
10 8 V
10 7 V
10 7 V
9 7 V
10 7 V
10 8 V
9 7 V
10 7 V
10 7 V
10 7 V
9 8 V
10 7 V
10 7 V
10 7 V
9 7 V
10 7 V
10 7 V
9 7 V
10 7 V
10 7 V
10 7 V
9 6 V
10 7 V
10 6 V
10 7 V
9 6 V
10 7 V
10 6 V
9 6 V
10 6 V
10 6 V
10 5 V
9 6 V
10 5 V
10 6 V
10 5 V
9 4 V
10 5 V
10 4 V
9 5 V
10 4 V
10 3 V
10 4 V
9 3 V
10 3 V
10 2 V
10 2 V
9 2 V
10 2 V
10 1 V
10 0 V
9 1 V
1.000 UL
LTb
3019 1131 M
1 -7 V
0 -7 V
1 -7 V
1 -7 V
1 -7 V
1 -8 V
1 -7 V
1 -7 V
1 -7 V
0 -7 V
1 -7 V
1 -7 V
1 -7 V
1 -7 V
0 -7 V
1 -7 V
1 -7 V
1 -7 V
1 -7 V
1 -7 V
1 -7 V
1 -7 V
0 -7 V
2 -7 V
0 -7 V
1 -7 V
1 -7 V
1 -7 V
1 -7 V
1 -7 V
0 -8 V
1 -7 V
1 -7 V
1 -7 V
1 -7 V
1 -7 V
0 -7 V
1 -7 V
1 -7 V
1 -7 V
1 -7 V
1 -7 V
0 -7 V
1 -7 V
1 -7 V
1 -7 V
1 -7 V
1 -7 V
0 -7 V
1 -7 V
1 -7 V
1 -7 V
0 -7 V
2 -7 V
0 -7 V
1 -8 V
1 -7 V
1 -7 V
1 -7 V
0 -7 V
1 -7 V
1 -7 V
1 -7 V
1 -7 V
0 -7 V
1 -7 V
1 -7 V
1 -7 V
1 -7 V
0 -7 V
1 -7 V
1 -7 V
1 -7 V
1 -7 V
1 -7 V
1 -7 V
0 -7 V
1 -7 V
1 -7 V
1 -7 V
1 -8 V
1 -7 V
0 -7 V
1 -7 V
1 -7 V
1 -7 V
0 -7 V
1 -7 V
1 -7 V
1 -7 V
1 -7 V
1 -7 V
0 -7 V
1 -7 V
1 -7 V
1 -7 V
0 -7 V
1 -7 V
1 -7 V
1 -7 V
1 -7 V
0 -7 V
1 -7 V
1 -7 V
1 -7 V
1 -8 V
1 -7 V
0 -7 V
1 -7 V
1 -7 V
1 -7 V
1 -7 V
0 -7 V
1 -7 V
1 -7 V
1 -7 V
1 -7 V
1.000 UL
LTb
2460 1834 M
6 0 V
5 0 V
6 0 V
6 -1 V
5 0 V
6 -1 V
6 0 V
5 -1 V
6 -1 V
5 -1 V
6 -1 V
6 -1 V
5 -1 V
6 -1 V
6 -1 V
5 -2 V
6 -1 V
6 -2 V
5 -2 V
6 -2 V
6 -2 V
5 -2 V
6 -3 V
6 -3 V
5 -2 V
6 -4 V
5 -3 V
6 -3 V
6 -4 V
5 -3 V
6 -4 V
6 -5 V
5 -4 V
6 -5 V
6 -4 V
5 -5 V
6 -6 V
6 -5 V
5 -6 V
6 -5 V
6 -6 V
5 -7 V
6 -6 V
5 -7 V
6 -6 V
6 -7 V
5 -8 V
6 -7 V
6 -8 V
5 -8 V
6 -8 V
6 -8 V
5 -9 V
6 -8 V
6 -9 V
5 -10 V
6 -9 V
6 -10 V
5 -10 V
6 -10 V
5 -10 V
6 -11 V
6 -11 V
5 -11 V
6 -12 V
6 -12 V
5 -12 V
6 -12 V
6 -13 V
5 -13 V
6 -13 V
6 -14 V
5 -14 V
6 -14 V
6 -14 V
5 -15 V
6 -15 V
5 -14 V
6 -15 V
6 -16 V
5 -15 V
6 -15 V
6 -15 V
5 -15 V
6 -15 V
6 -14 V
5 -14 V
6 -14 V
6 -12 V
5 -12 V
6 -11 V
6 -10 V
5 -9 V
6 -7 V
5 -5 V
6 -3 V
6 -1 V
5 1 V
6 5 V
1.000 UL
LTb
2182 1834 M
3 0 V
3 0 V
3 0 V
2 0 V
3 0 V
3 0 V
3 0 V
3 0 V
2 0 V
3 0 V
3 0 V
3 0 V
3 0 V
3 0 V
2 0 V
3 0 V
3 0 V
3 0 V
3 0 V
2 0 V
3 0 V
3 0 V
3 0 V
3 0 V
2 0 V
3 0 V
3 0 V
3 0 V
3 0 V
2 0 V
3 0 V
3 0 V
3 0 V
3 0 V
2 0 V
3 0 V
3 0 V
3 0 V
3 0 V
2 0 V
3 0 V
3 0 V
3 0 V
3 0 V
3 0 V
2 0 V
3 0 V
3 0 V
3 0 V
3 0 V
2 0 V
3 0 V
3 0 V
3 0 V
3 0 V
2 0 V
3 0 V
3 0 V
3 0 V
3 0 V
2 0 V
3 0 V
3 0 V
3 0 V
3 0 V
2 0 V
3 0 V
3 0 V
3 0 V
3 0 V
2 0 V
3 0 V
3 0 V
3 0 V
3 0 V
2 0 V
3 0 V
3 0 V
3 0 V
3 0 V
3 0 V
2 0 V
3 0 V
3 0 V
3 0 V
3 0 V
2 0 V
3 0 V
3 0 V
3 0 V
3 0 V
2 0 V
3 0 V
3 0 V
3 0 V
3 0 V
2 0 V
3 0 V
3 0 V
1.000 UL
LTa
500 1134 M
29 0 V
30 0 V
29 0 V
30 0 V
29 0 V
30 0 V
29 0 V
29 0 V
30 0 V
29 0 V
30 0 V
29 0 V
30 0 V
29 0 V
29 0 V
30 0 V
29 0 V
30 0 V
29 0 V
30 0 V
29 0 V
29 0 V
30 0 V
29 0 V
30 0 V
29 0 V
29 0 V
30 0 V
29 0 V
30 0 V
29 0 V
30 0 V
29 0 V
29 0 V
30 0 V
29 0 V
30 0 V
29 0 V
30 0 V
29 0 V
29 0 V
30 0 V
29 0 V
30 0 V
29 0 V
30 0 V
29 0 V
29 0 V
30 0 V
29 0 V
30 0 V
29 0 V
30 0 V
29 0 V
29 0 V
30 0 V
29 0 V
30 0 V
29 0 V
30 0 V
29 0 V
29 0 V
30 0 V
29 0 V
30 0 V
29 0 V
30 0 V
29 0 V
29 0 V
30 0 V
29 0 V
30 0 V
29 0 V
29 0 V
30 0 V
29 0 V
30 0 V
29 0 V
30 0 V
29 0 V
29 0 V
30 0 V
29 0 V
30 0 V
29 0 V
30 0 V
29 0 V
29 0 V
30 0 V
29 0 V
30 0 V
29 0 V
30 0 V
29 0 V
29 0 V
30 0 V
29 0 V
30 0 V
29 0 V
1.000 UL
LTa
500 1834 M
29 0 V
30 0 V
29 0 V
30 0 V
29 0 V
30 0 V
29 0 V
29 0 V
30 0 V
29 0 V
30 0 V
29 0 V
30 0 V
29 0 V
29 0 V
30 0 V
29 0 V
30 0 V
29 0 V
30 0 V
29 0 V
29 0 V
30 0 V
29 0 V
30 0 V
29 0 V
29 0 V
30 0 V
29 0 V
30 0 V
29 0 V
30 0 V
29 0 V
29 0 V
30 0 V
29 0 V
30 0 V
29 0 V
30 0 V
29 0 V
29 0 V
30 0 V
29 0 V
30 0 V
29 0 V
30 0 V
29 0 V
29 0 V
30 0 V
29 0 V
30 0 V
29 0 V
30 0 V
29 0 V
29 0 V
30 0 V
29 0 V
30 0 V
29 0 V
30 0 V
29 0 V
29 0 V
30 0 V
29 0 V
30 0 V
29 0 V
30 0 V
29 0 V
29 0 V
30 0 V
29 0 V
30 0 V
29 0 V
29 0 V
30 0 V
29 0 V
30 0 V
29 0 V
30 0 V
29 0 V
29 0 V
30 0 V
29 0 V
30 0 V
29 0 V
30 0 V
29 0 V
29 0 V
30 0 V
29 0 V
30 0 V
29 0 V
30 0 V
29 0 V
29 0 V
30 0 V
29 0 V
30 0 V
29 0 V
stroke
grestore
end
showpage
}}%
\put(352,1835){\makebox(0,0)[l]{$\alpha_c$}}%
\put(352,1131){\makebox(0,0)[l]{$\alpha_d$}}%
\put(1127,1638){\makebox(0,0)[l]{$x$-UNSAT}}%
\put(1606,652){\makebox(0,0)[l]{$x$-SAT}}%
\put(1975,50){\makebox(0,0){$x$}}%
\put(100,1180){%
\special{ps: gsave currentpoint currentpoint translate
270 rotate neg exch neg exch translate}%
\makebox(0,0)[b]{\shortstack{$\alpha$}}%
\special{ps: currentpoint grestore moveto}%
}%
\put(3450,200){\makebox(0,0){ 0.8}}%
\put(3081,200){\makebox(0,0){ 0.7}}%
\put(2712,200){\makebox(0,0){ 0.6}}%
\put(2344,200){\makebox(0,0){ 0.5}}%
\put(1975,200){\makebox(0,0){ 0.4}}%
\put(1606,200){\makebox(0,0){ 0.3}}%
\put(1238,200){\makebox(0,0){ 0.2}}%
\put(869,200){\makebox(0,0){ 0.1}}%
\put(500,200){\makebox(0,0){ 0}}%
\put(450,2060){\makebox(0,0)[r]{ 0.95}}%
\put(450,1708){\makebox(0,0)[r]{ 0.9}}%
\put(450,1356){\makebox(0,0)[r]{ 0.85}}%
\put(450,1004){\makebox(0,0)[r]{ 0.8}}%
\put(450,652){\makebox(0,0)[r]{ 0.75}}%
\put(450,300){\makebox(0,0)[r]{ 0.7}}%
\end{picture}%
\endgroup
 